\pdfoutput=1
\documentclass{aa}

\usepackage{txfonts}

\usepackage{graphicx}
\usepackage{color}
\usepackage{url}

\usepackage[]{natbib}
\bibpunct{(}{)}{;}{a}{}{,} 

\newcommand{\referee}[1]{{#1}}

\newcommand{\myfigure}[3]{
  \begin{figure}
    \resizebox{\hsize}{!}{\includegraphics{#1}}
    \caption{#2}
    \label{#3}
  \end{figure}
}

\begin{document}

\title{Probing the nuclear star cluster of galaxies with ELTs} 

\date{\today} 

\author{
  M. Gullieuszik\inst{1}\and
  L. Greggio\inst{1}\and
  R. Falomo\inst{1}\and
  L. Schreiber\inst{2}\and
  M. Uslenghi\inst{3}  
}
\institute{
  INAF, Osservatorio Astronomico di Padova, 
  Vicolo dell'Osservatorio 5, I-35122 Padova, Italy
  \and
  INAF, Osservatorio Astronomico di Bologna, 
  Via Ranzani 1, I-40127 Bologna, Italy
  \and
  INAF, Istituto di Astrofisica Spaziale e Fisica Cosmica, 
  Via Bassini 15, I-20133 Milano, Italy 
  }

\abstract{
      The unprecedented sensitivity and spatial resolution of next
      generation ground-based extremely large telescopes (ELTs) will
      open a completely new window on the study of resolved stellar
      populations.
      In this paper we study the feasibility of the analysis of
      nuclear star cluster (NSC) stellar populations with ELTs.  To
      date, NSC stellar population studies are based on the properties
      of their integrated light. NSC are in fact observed as
      unresolved sources even with the HST. We explore the
      possibility to obtain direct estimates of the age of NSC stellar
      populations from photometry of main-sequence
      turn-off stars.
We simulated ELT observations of NSCs at different distances and with
different stellar populations. Photometric measurements on each
simulated image \referee{were} analysed in detail and results about photometric
accuracy and completeness are reported \referee{here}.
We found that the main-sequence turn-off is detectable --and therefore
the age of stellar populations can be directly estimated-- up to 2Mpc
for old, up to 3 Mpc for intermediate-age and up to 4-5 Mpc for young
stellar populations.  We found that for this particular science case
the performances of TMT and E-ELT are of comparable quality.}
\maketitle

\section{Introduction}\label{sec:intro}

The nuclei of galaxies are extraordinary laboratories to probe the
processes that lead to the formation of nowadays galaxies.  The
properties of the nuclear regions are in fact thought to be linked
with the formation history of the galaxies where collapse of gas and
merging events may play a relevant role for different type of galaxies
and environments.  Various phenomena characterize these nuclei such as
super-massive black holes (SMBHs), active galactic nuclei,
central starbursts and extreme stellar densities. All these phenomena
are likely connected to the global properties of their host galaxies
and co-evolution of the various components are invoked in order to
account for the observed correlations.

For instance important links between the mass of the central SMBH and
the global properties of the host galaxies have been discovered
\citep[see e.g.][and refs
  therein]{ferr+2006,wehn+2006,neum+2012,scot+2013}.  Another less
explored phenomenon is the presence of compact and massive star
clusters found in the centres of most of nearby galaxies of all
morphological types \citep[see e.g.][]{boke+2010}.  The formation of
nuclear star clusters (NSC) may also be connected to the formation of
SMBH in the centre and these objects could therefore
trace the evolution history of the whole galaxy formation. For
example, NSCs sit on the low mass extrapolation of the scaling
relation between the SMBH mass and the total mass of the host galaxy.

\referee{Photometry of NSCs in a large sample of
 spiral galaxies was investigated by \cite{boke+2002} and,
 more recently, by \citet{geor+2014}. NSCs in spheroidal
 galaxies were studied by \cite{cote+2006} and \cite{turn+2012}.
}
NSCs are 3-4 mag \citep{geor+2014} more luminous than Milky Way
globular clusters,  but they share  similar sizes \citep[few
parsecs, see e.g.][]{boke+2004}. Their estimated  total mass \citep[$10^6$-$10^7$ $M_\odot$;][]{boke+2010} is 
at the high end of the globular clusters' mass function.

Two main  scenarios have been suggested for the formation of NSCs: 
i) infall of star clusters that formed elsewhere in the host galaxy \citep{trem+1975,capu+2008} 
ii) {\it in situ} formation and build-up through star formation following the accretion of gas in the centre of the galaxies.
performed chemodynamical simulations of the inner regions of dwarf
galaxies, where dissipative merging of stellar and gaseous clumps
formed in the spiral arms can lead to the formation of NSCs.
According to this model the metallicity of the stars in the NSC would
be higher than that of its host spheroid. In addition, in these models
the times-scale of cluster formation is shorter for higher masses \referee{and}
therefore one expects that less massive galaxies should host younger
NSCs.

A detailed study of the stellar populations in NSCs appears therefore
a key tool to understand what are the main physical processes involved
in the formation and growth of these stellar systems.  However,
current instrumentation allows us to investigate this issue only from
observations of the NSCs integrated light. Ages, metallicities and in
general star formation histories are inferred from integrated colours,
indices, and/or spectral energy distribution fitting techniques.
\citep[see e.g.,][]{walc+2006, seth+2006,ross+2006,seth+2010}.  In
fact even with the HST capabilities it is not possible to resolve the
stars of NSCs beyond the Local Group \cite[see
  e.g.][]{boke+2002}\footnote{The typical size of a NSC ($\sim$ 3 pc)
  corresponds to $0\farcs3$ at 2 Mpc}. The derived properties of the stellar
population are hence based on population synthesis models and results
are therefore affected by uncertainties intrinsically related to the
models, like the age-metallicity degeneracy \referee{as well as} the difficulty of
separating different age components. In particular the presence of
young stars strongly limits the accuracy of the age and metallicity
estimate of unresolved stellar systems because the young component dominates the
integrated light, even when representing a small contribution to the
mass of the whole stellar system.

Recent studies from HST spectroscopic observations \citep{paud+2011}  of 
early-type dwarf galaxies in the Virgo cluster  
suggest that many nuclei contain a stellar population that is 
significantly younger (by $\sim$ 3 Gyr ) and more metal-rich than that of the 
host galaxy. This would indicate that NSCs formed after the main body of the host galaxy,
likely from mergers of gas clumps. 
However in a small number of cases the nuclei are found to be dominated by 
older and more metal poor stellar population than that of their host galaxies, thus favouring the 
merging scenario of old pre-existing star clusters.
These differences could be reconciled if different mechanisms were invoked for NSCs in galaxies of 
different mass. For example, from the study of  the 
surface brightness profiles of a sample of $\sim$ 40 early-type galaxies in the Fornax cluster, \cite{turn+2012}
suggest  that the formation of NSCs occurs   via merging of pre-existing clusters in lower mass galaxies, via gas
  accretion in higher mass hosts.

The emerging picture from these studies is limited by the use of the integrated light of the stellar systems 
and significant advancement in recovering the star formation history of the nuclei of these  galaxies 
can be derived from the photometry of individual stars. 
The analysis of the colour-magnitude diagram (CMD) of the resolved stars is in fact the most direct way to estimate the age and the metallicity
of stellar populations. 
Indeed, isochrone fitting of  the turn off region is the optimal tool to derive the age, while the colour of the
  red giant branch stars is a good estimator of the metallicity 
 \citep[see e.g.][]{greg+2011}. In addition, resolving individual stars  would yield a direct information on the multiplicity of star
 formation events, or on the duration of the episodes of star formation in the  NSCs. 

Given the high surface density of stars in NSCs and the large distance
of their host galaxies, the construction of CMDs of stars is well
beyond the capabilities of present-day instrumentation but will become
feasible with the next generation of extremely large telescopes (ELTs)
--as the European-ELT (E-ELT)\footnote{\url{http://www.eso.org/sci/facilities/eelt/}},
the Thirty Meter Telescope (TMT)\footnote{\url{http://www.tmt.org/}},
the Grand Magellan Telescope (GMT)\footnote{\url{http://www.gmto.org/}}-- that will join high sensitivity with
extraordinary (few mas) spatial resolution, when coupled with adaptive
optics.  In particular, the unprecedented spatial resolution of
Adaptive-Optics ELTs (3-5 mas) will open a new window in the study of
crowded stellar fields. It will be possible, for the first time, to
obtain resolved stellar photometry in NSCs beyond the Local Group
yielding direct measurements of stellar ages and metallicities.  We
will then be in the condition to provide strong constraints to the
physical models of NSC formation and to understand what are the
connections between the formation and evolution of central massive
objects and their host galaxy.

In this paper we explore in detail the capabilities of studying the
stellar population of NSCs in nearby galaxies (up to 4 Mpc) using simulated
near-infrared (IR) imaging observations of future ELTs. In particular we adopt as
baseline the parameters of Multi-AO Imaging Camera for Deep
Observations \citep[MICADO;][]{davi+2010} proposed for E-ELT and
compare the results to those obtainable with the InfraRed Imaging
Spectrograph \citep[IRIS;][]{lark+2010} foreseen for the TMT. The results are also
compared with the expected capabilities of the Near Infrared Camera
\citep[NIRCam;][]{gree+2010} to be mounted at the James Webb Space
telescope (JWST) that may reach similar sensitivity but with worse
angular resolution.

The paper is organised as follows:
Section \ref{sec:sim} describes our simulated observations of NSCs.
Section \ref{sec:phot} describes the procedure we adopted to extract
photometric catalogues from the simulated images and the analysis of the
photometric accuracy.  Section \ref{sec:instruments} presents an
analysis of the feasibility of the science case with other
next-generation instrumentation.  A summary of this work and our
conclusions are presented in Section \ref{sec:conclusions}.


\section{Simulation of NSCs in galaxies}\label{sec:sim}

In order to age-date the resolved stellar population of
the NSC  we need to disentangle the cluster members from the stars 
of its host galaxy.  In the central parts of the
cluster, the stellar field is dominated by the cluster members, but
the stellar density is very high and stars are blended, decreasing the
accuracy of the photometry.  On the other hand, in the external regions, where
photometry is more accurate, the host galaxy \referee{body} population contribution
becomes important, and it is more difficult to disentangle the two
components. The interplay of these two factors determines the
intermediate region where photometry can be accurate enough and the host
contribution is low enough to allow us to measure reliable age and metallicity
for the cluster stellar populations.  Thus, the possibility of
characterizing the stellar population of NSCs depends on a few
parameters, e.g. the distance of the host galaxy and the
core radius of the cluster which control the crowding conditions. In
addition, for a given distance, the older the NSC the fainter are the stars at the turn
off,  which makes it more difficult to determine its age.
In order to exploit the capabilities of ELTs to investigate NSCs, we have considered a number of cases with different age and at different distance. 

\subsection{The input stellar populations}

As baseline for the simulations we adopted NGC~300 as a NSC-host
template galaxy.  This late-type galaxy located at 2 Mpc is one of the closest galaxies known to host a NSC.
\cite{boke+2004} found that the NSC in NGC~300 follows a King profile
with a tidal radius $r_t=2\farcs85$ and core radius $r_c= 0\farcs095$.
The resulting half-light radius is $r_e=0\farcs27$. 
\referee{Using a single age and single metallicity stellar population (Simple Stellar Population, SSP) model, the age of the
  stellar population in the NSC is estimated to be $\sim 1$ Gyr.}
Allowing for an extended star formation period, the data support a luminosity-weighted mean age of 3 Gyr, and a wide range in age. 
The NSC metallicity is estimated of $Z=0.004$, and the total mass of 
$M=10^6M_\sun$ 
\citep{ross+2006,walc+2006}.

In the simulated frames, 
the input stellar population consists of a combination of two
components, one for the host galaxy \referee{body}, described as a composite
stellar population, and one for the NSC, described as a SSP.
While there is no guarantee \referee{that} the NSCs are really SSPs, our simulations will show how well the simplest case of NSC stellar population can be characterized in terms of age and metallicity. More complicated star formation histories can be viewed as the superposition of several individual episodes of star formation.

For the host galaxy \referee{body} stellar population we adopted the model
used in \cite{greg+2012} as representative of a typical disk of a late-type
galaxy. The model has a constant star formation rate over the last 12 Gyr, and a
chemical evolution with the metallicity linearly increasing from $Z=0$
up to the solar value in the first 5 Gyrs. Later on, 
the chemical enrichment is assumed much slower, with the metallicity 
reaching $Z=0.02$ for the youngest stars. 
\referee{The surface density of stars for the host-galaxy body component was chosen
so as to reproduce a central surface brightness of 
$\mu_B= 20.4$ mag arcsec$^{-2}$. For the adopted
galaxy model, this corresponds to surface brightness of 18.88,
17.98, 17.27, and 17.13 mag arcsec$^{-2}$ in the $I$, $J$, $H$, and $K$ bands,
respectively. These values are close to those measured for NGC~300
\citep[see e.g.][]{boke+2002}.}

\begin{table}
  \caption{Parameters of the SSPs used to simulate the NSCs:
  	age, metallicity, total number of stars, $I$-band absolute magnitude and total mass at birth.}
  \label{tab:sim}  
  \centering
  \begin{tabular}{ccccc}
    \hline\hline
    $\tau$          & $Z$ & $N$ & $M_I$& $M$\\
    $[$Gyr$]$ &     &     &      & $[10^6M_\sun]$\\
    \hline
1 &  0.019 & $4\cdot10^5$& -11.3 &1.1 \\ 
4 &  0.019 & $4\cdot10^5$& -10.3 &1.2 \\ 
10&  0.004 & $5\cdot10^5$& -10.4 &2.0 \\
    \hline      
  \end{tabular}
\end{table}

We choose an SSP model to describe the stellar population of
the NSC, for which we considered three age values, 1, 4, and
10 Gyr. This allows us to probe the dependence of the results on the
NSC age.
The synthetic SSP was created by distributing objects along 
stellar isochrones from \cite{mari+2008} provided by the CMD
v2.5 web tool\footnote{\url{http://stev.oapd.inaf.it/cgi-bin/cmd}}.
We adopted a Salpeter initial mass function from $0.5 M_\sun$ to the
mass of the star at the tip of the Asymptotic Giant Branch (AGB). The contribution to the total
luminosity of stars with masses below $0.5 M_\sun$ is negligible
($\lesssim 5 \%$ for all considered cases in all considered bands, assuming a flat initial mass function below 0.5 $M_\sun$). The main properties
of the input SSPs used to simulate the NSCs are summarised   in Table
\ref{tab:sim}.

The dependence of the results on the galaxy distance was tested by
putting our synthetic populations at 2 and 4 Mpcs.  The simulated NSC
stars have been distributed following a King-profile model; for the
simulations of the systems at 2 Mpc we assumed the structural
parameters of the NSC in NGC~300 (see above), while for those at 4~Mpc
we adopted $r_t=1\farcs43$ and $r_e=0\farcs13$, consistently scaled
from those of the NSC in  NGC~300.

\subsection{Simulated images}

\begin{figure*}
  \centering
  \includegraphics[width=18cm]{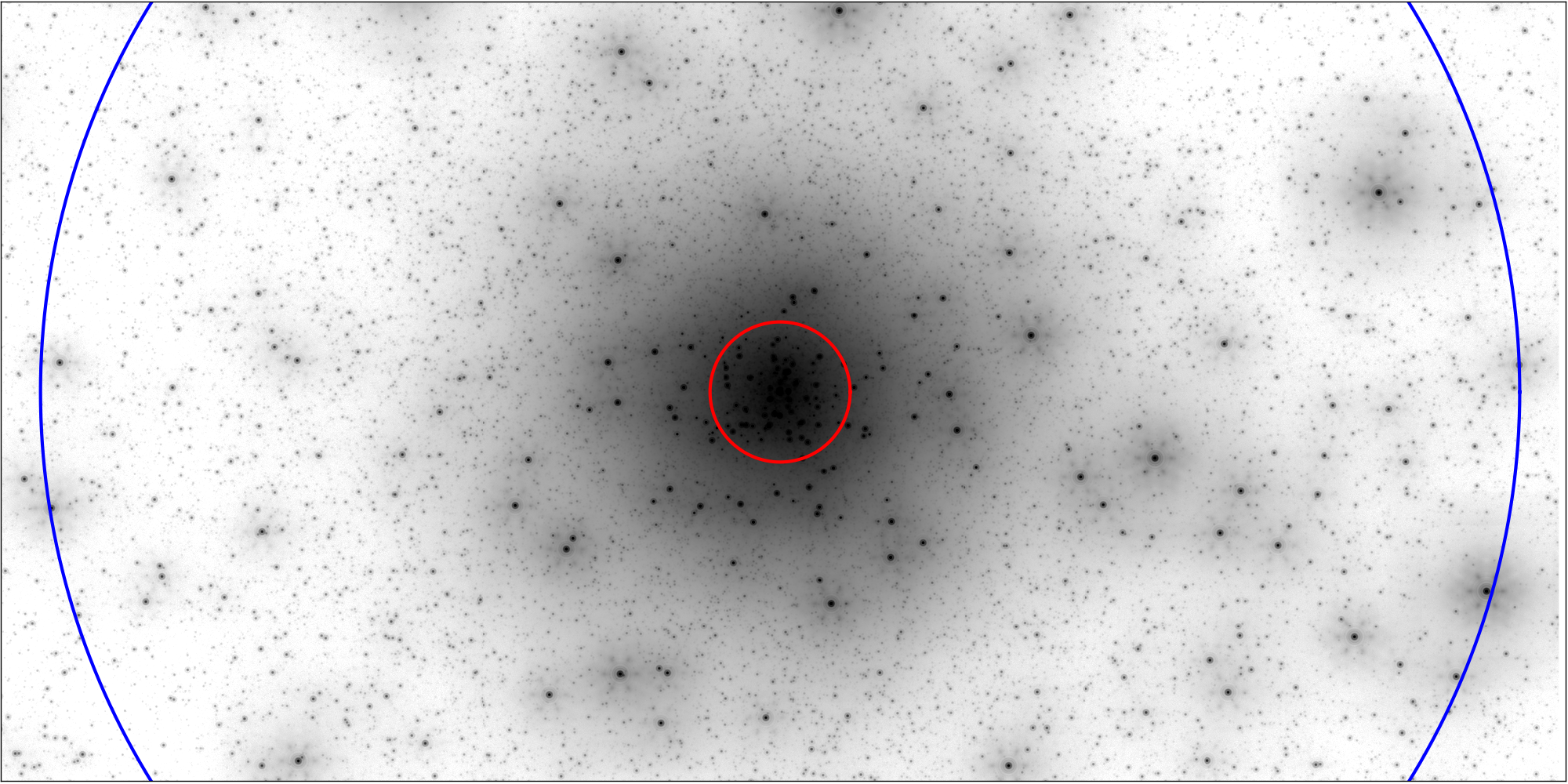}\\
  \caption{Zoom of a $6\arcsec\times3\arcsec$ region of the $J$-band
    simulated \referee{MICADO@E-ELT} image of the 1 Gyr old NSC located at 2 Mpc.  The
    circles shows the effective radius ($0\farcs27$, in {\it red}) and the tidal radius ($2\farcs85$, in {\it blue}) of the
    cluster.}
  \label{fig:ima}
\end{figure*}

\referee{

We simulated observations with two instruments,
  \mbox{MICADO@E-ELT} and \mbox{IRIS@TMT} following the 
  specifications in \cite{davi+2010} and \cite{wrig+2010}.}

\referee{MICADO (Multi-AO Imaging Camera for Deep Observations) is a
  phase A imaging instrument designed for E-ELT to provide diffraction
  limited imaging over a wide ($\sim 1$ arcmin) field of view.}  The
simulations of MICADO observations \referee{in $I$, $J$, $H$, and
  $K$-bands} were performed assuming a primary mirror with a diameter
of 39 meters and a 28\% obscuration.  We assumed a total throughput of
39\% in the $I$ and $K$ bands and 40\% in $J$ and $H$ bands.  We used
a read noise of 5 $e^-$, and a pixel scale 3 mas.  We assumed the
typical near-IR sky background measured at Cerro Paranal and included
the contribution of thermal emission in the near-IR-bands;
specifically we adopted a total background of 20.0, 16.3, 15.0, and
12.8 mags/arcsec$^{-2}$ in the $I$, $J$, $H$, and $K$ bands,
respectively.  \referee{Under these conditions the signal-to-noise of
  isolated point sources of magnitude 26-28 is dominated by the
  contribution of the background while read-out noise is
  negligible. This allows one to obtain several individual frames of
  the same field in order to avoid saturation and properly subtract
  the bright IR background.}  We used the last version (June 2012) of
the PSFs of the Multi Conjugate Adaptive Optics post focal relay
\cite[MAORY;][]{diol+2010} calculated for a $0\farcs6$ seeing at the
centre of the corrected field of view (FoV), as provided by the MAORY
official
website\footnote{\url{http://www.bo.astro.it/maory/Maory/Welcome.html}}.

The InfraRed Imaging Spectrograph \citep[IRIS;][]{lark+2010} is one of
the first-light instruments designed for the TMT.  
\referee{TMT is a 30-m telescope that will be mounted at Mauna Kea, Hawaii, USA.
It will observe in the wavelength range from 0.31 to 20 $\mu$m.}
IRIS will have both
imaging and integral field spectroscopy capabilities with a spectral
coverage from $0.84$ $\mu m$ to $2.4$ $\mu m$.  The imager has 4 mas
pixels and a field of view of $16\farcs4$ and will be equipped
with $Z$, $Y$, $J$, $H$, and $K$ filters.  A full description of the
IRIS instrument and its performances can be found in \cite{wrig+2010} and \cite{do+2014}.
We simulated IRIS observations using the input star list (magnitudes
and positions on the frame) generated for the MICADO images for the 1
Gyr old SSP at 2 Mpc distance (details in Sect. \ref{sec:instruments}). Images were produced
using the PSFs kindly provided by S. Wright and the NFIRAOS Team
\cite[see][]{do+2014}.
\referee{TMT observations have been simulated in the $Z$ and $J$ band only.
}

All simulated images were produced using the
AETC\footnote{\url{http://aetc.oapd.inaf.it/}} v.3.0 tool
\citep{falo+2011} using a total integration time of 3$h$.  \referee{We
  simulated a simple observing strategy, with no dithering because the
  PSF core is well sampled in $J$, $H$ and $K$ bands and slightly
  undersampled in the $I$ band (the FWHM of the PSF core is 1.8
  pixels). The impact of this on the $I$-band photometric accuracy is not a
  major issue (see Sect. \ref{sec:photerr}).}  The size of the images
was set much larger than the tidal radius of the NSC.  For simulations
at 2 Mpc we produced \referee{
images with a $15\farcs0$ FoV;
at 4 Mpc we reduced the size to 
FoV=$7\farcs5$.}
  An example of
a simulated image is shown in Fig.~\ref{fig:ima}.

\section{Photometry on E-ELT images}\label{sec:phot}

The photometric analysis of the simulated images was performed using
Starfinder \citep{diol+2000}, a photometric package specifically
designed for high resolution AO images.  One of the key-features of
Starfinder is that it creates a 2D image of the PSF using bright and
isolated stars, without any analytic approximation. This is
particularly important since the PSF in AO systems is highly
structured; therefore it can not be accurately
described as a combination of few analytical components
\citep[see for details][]{schr+2012}.  Starfinder was already tested and proved to
provide excellent results on E-ELT simulated images
\citep{deep+2011,schr+2014}. 

The photometric analysis was performed assuming we had no {\it a
  priori} knowledge of the PSF, simulating the photometric reduction
of real observations.  
We set a detection limit of $3\sigma$ above
the local background as the threshold for object detection.  
Firstly,
we masked out the crowded central region of the NSC in the image.
Then, a rough estimate of the background level was
obtained as the median value of the image and the PSF was estimated
from 50 bright and relatively isolated stars.
The background level and the PSF estimates were then refined
using an iterative procedure.
Each step consists of the following four operations:
\begin{itemize}
\item evaluation of the average value of the
background and its subtraction  from the image; 
\item estimation of a new PSF from 50 bright and isolated stars
after subtracting all neighbouring stars ;
\item fitting of all sources detected  at $3\sigma$ level above
the local background with the PSF model;
\item calculation of a residual image obtained by subtracting 
all fitted stars from the original
image; this image was then used to obtain an improved estimate of the background.
\end{itemize}
This procedure was iterated until the background value and the
PSF model converge to the final values, i.e. the variations with
respect to the previous iteration were less than 1\% (in all cases this
happened after four iteration).
The final source detection and fitting was performed on the original
image, including the central region previously masked out.

The photometric catalogues were  cross-correlated with the input
catalogues with a matching radius of 1 pixel.
As a first step in our matching procedure we sorted the photometric
catalogue by increasing magnitudes (from bright to faint objects).
When more than one input star was found within 1 pixel from a measured
source, the brightest input star was chosen. The matched input star was removed from the
catalogue and the procedure proceeded with the next star.
The zero-point of the
photometry was derived as the median difference between the input and
the measured magnitudes of the 300 brightest stars located
beyond the tidal radius of the NSC.  This was done because in the
crowded central regions the stars are systematically affected by a
zero-point offset due to blending of two or more sources. 
This issue
will be addressed in the next sections.

 \begin{figure*}[!ht]
    \centering
    \includegraphics[width=9.cm]{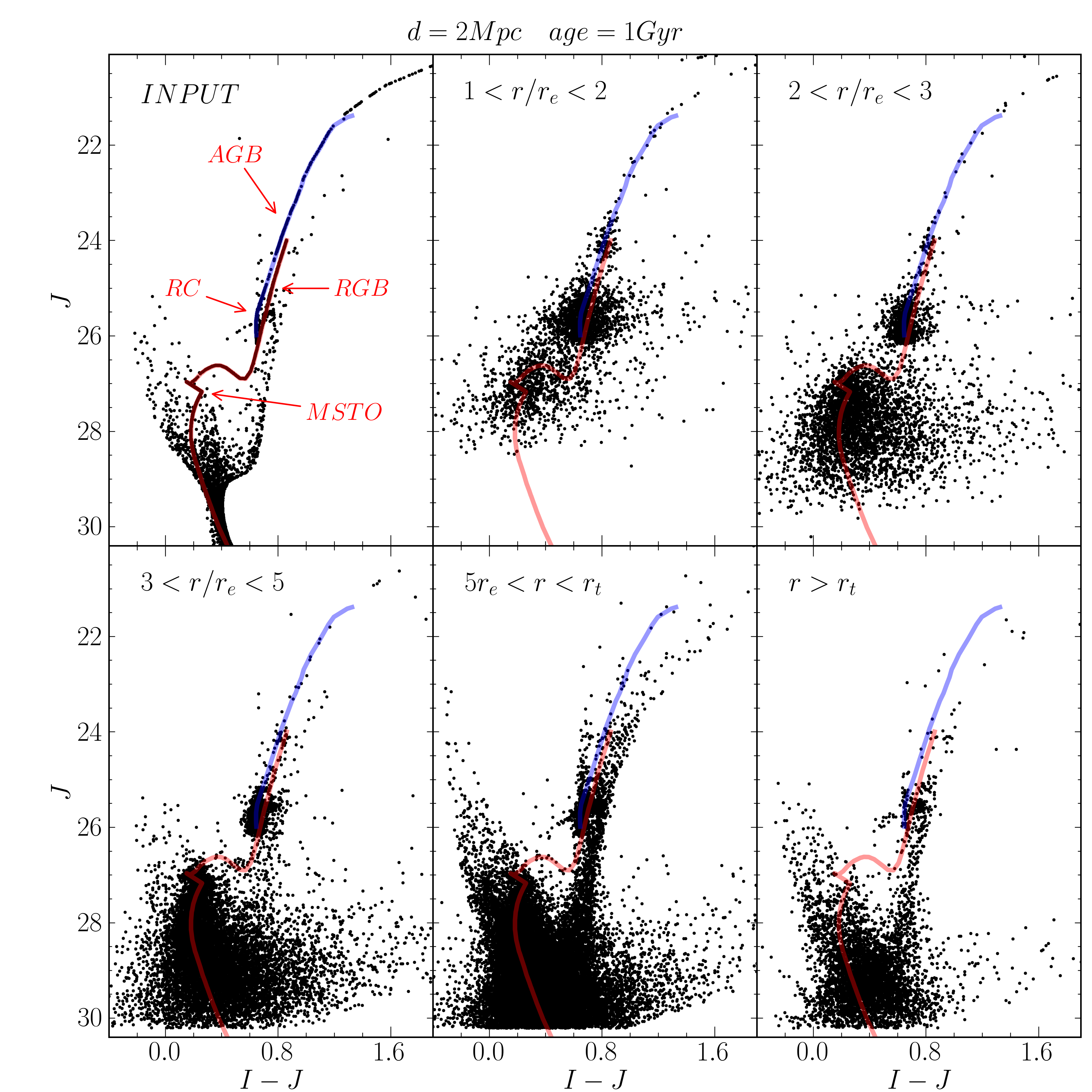}\hfill
    \includegraphics[width=9.cm]{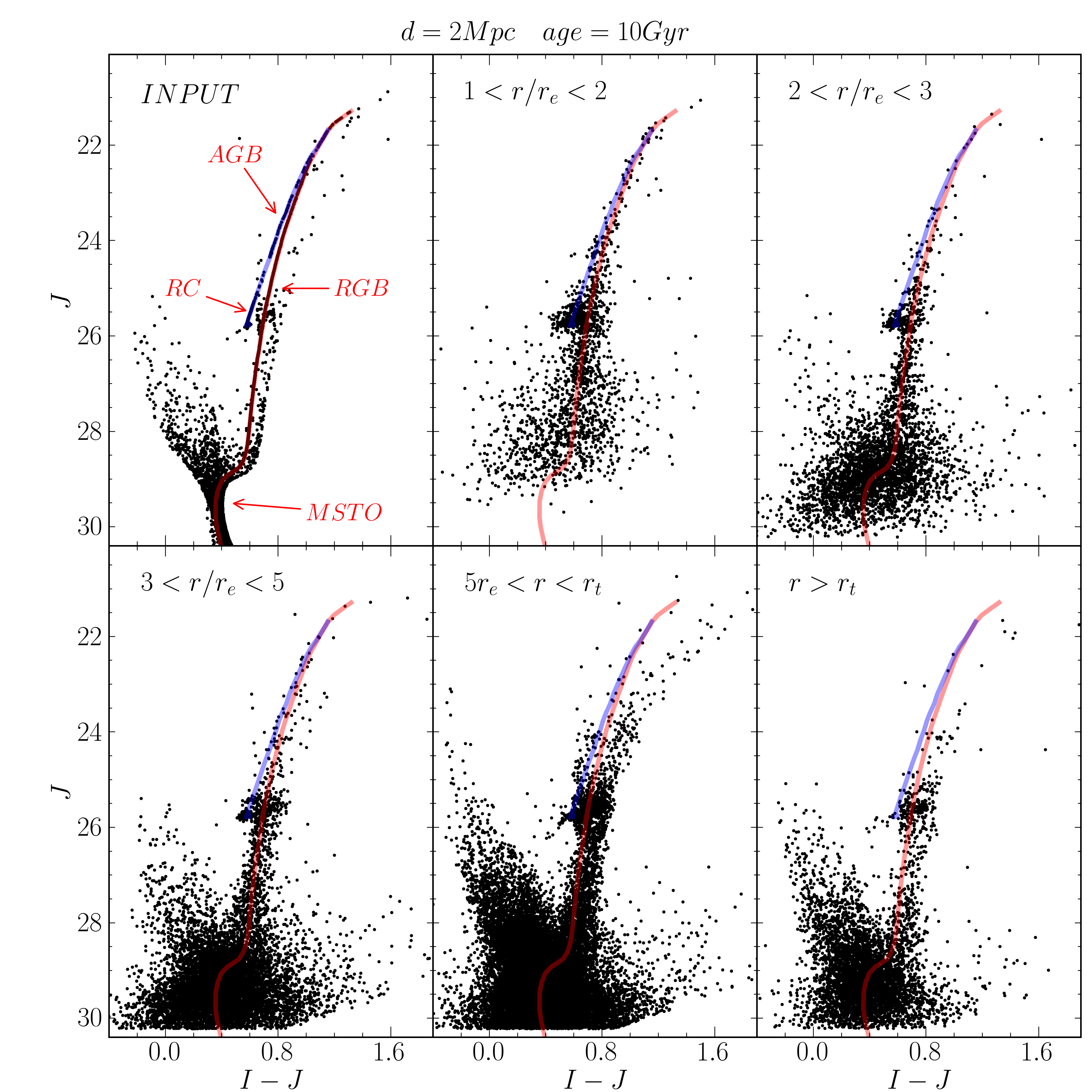}\\[2em]
    \includegraphics[width=9.cm]{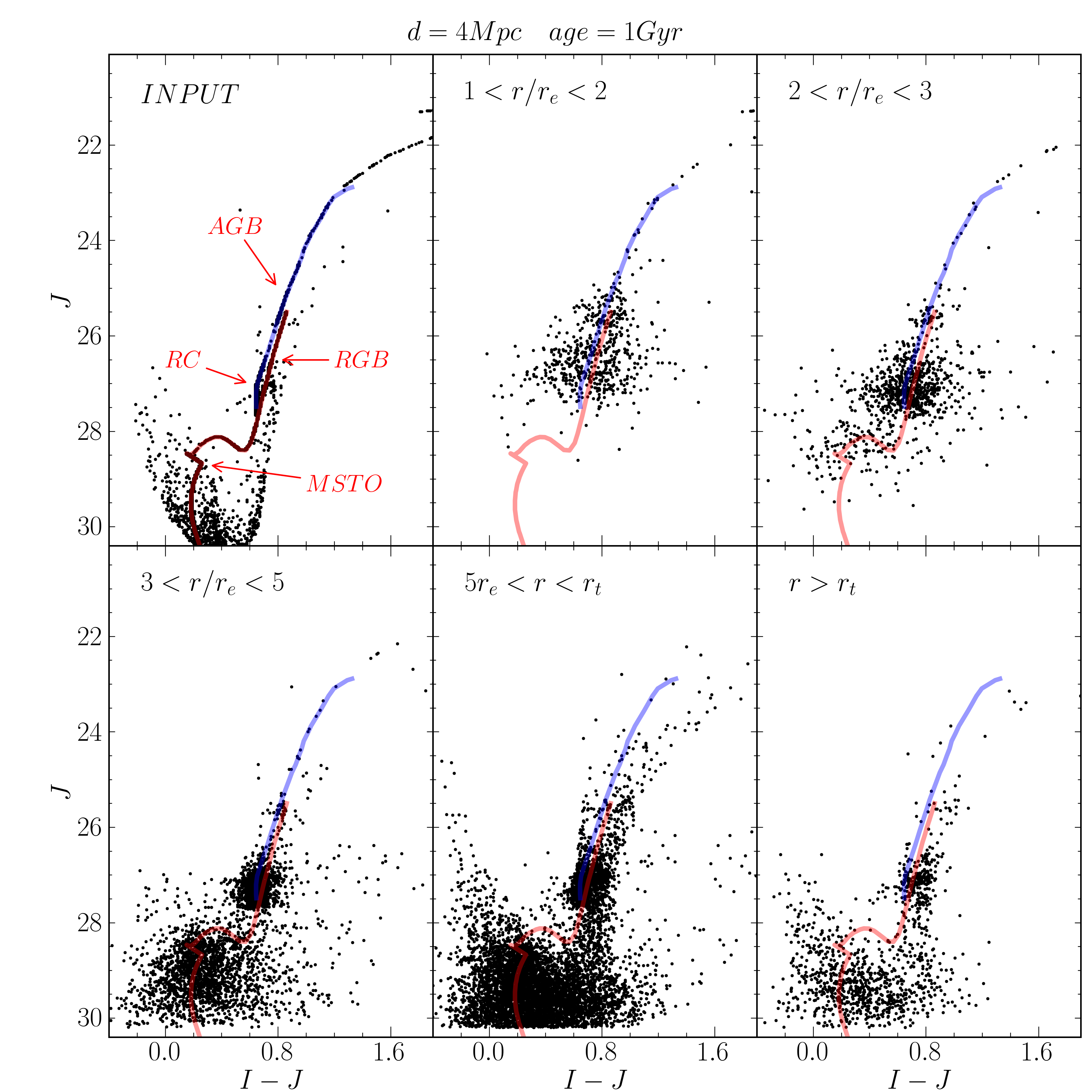}\hfill
    \includegraphics[width=9.cm]{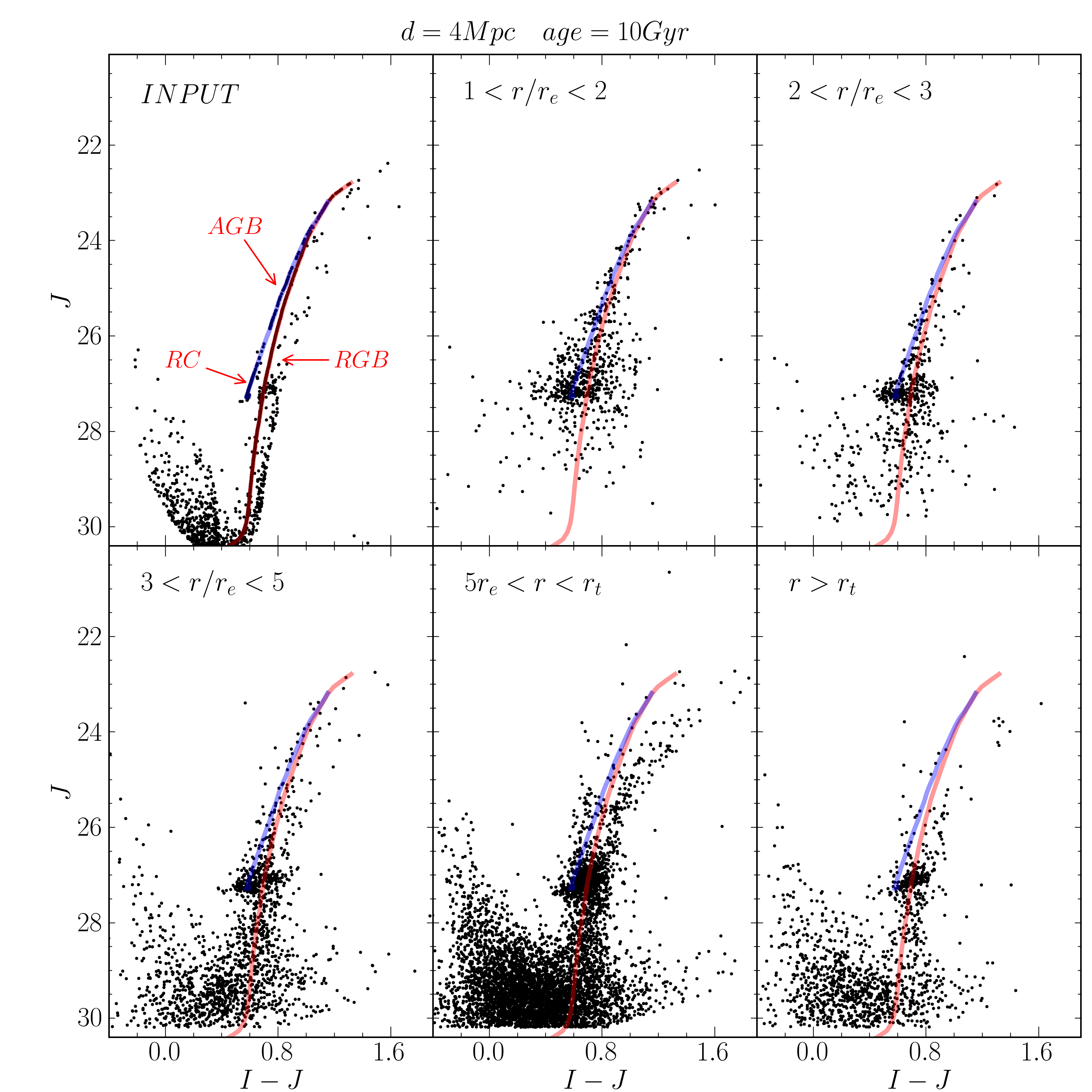}
    \caption{Colour-magnitude diagrams of different regions of the
      simulated stellar system \referee{obtained from MICADO simulations}.
      The lines are the isochrones from
      \cite{mari+2008} representative of the nuclear cluster stars.
      The {\it red line} is the MS and the RGB, while the {\it blue line}
      is the sequence of core He-burning stars, up to the first
      thermal pulse on the AGB.
      \referee{The upper-left sub-panels shows the CMDs
        of the input stellar catalogues (NSC plus host galaxy main body).
      For clarity only stars in the inner $3\,r_e$ region are shown.}
      }
    \label{fig:cmds}
  \end{figure*}

Some examples of CMDs obtained from the simulated images are shown in
Fig. \ref{fig:cmds} where the solid lines in each panel show the locus
described by the input stellar population.  \referee{The CMDs of
the input stellar catalogues are also shown for reference.}
In the crowded central regions of the clusters stellar blends strongly
decrease the photometric accuracy.  Photometric errors are extremely
large and the photometry is much shallower than in the outer
regions. As a consequence the CMDs in the central region can not be
used to study the NSC stellar populations.  Going towards larger
radii, the CMD becomes progressively more populated with the
underlying galaxy component, until the CMDs in the external regions
are mainly populated by host galaxy members, as the surface density of
NSC stars is very low.  The CMDs of the regions at $5r_e<r<r_t$ are in
fact undistinguishable from those of the regions beyond the NSC tidal
radius, populated only by members of the host galaxy.  The NSC stellar
populations are clearly visible in the intermediate regions, between 3
and 5 $r_e$.  This is especially true for the cases with a 1 Gyr old
NSC, at both 2 and 4 Mpc distance.  The brightest part of the
red-giant branch (RGB) and the red clump (RC) are visible in all CMDs.
The main-sequence turn-off (MSTO) is clearly visible in the most
favourable cases (e.g. 1 Gyr old SSP at 2 Mpc distance), but
completely undetectable for the oldest SSP at 4 Mpc. In the
intermediate cases, detection of the NSC MSTO is complicated by the
presence of the host galaxy population.  The main goal of this paper
is to study under which conditions it is possible to observe NSC MSTO
stars. To this end, it is necessary to assess whether the
contamination by foreground host galaxy stellar populations prevents
the detection or not.  We discuss this issue in
Sect. \ref{sec:stelpops}

\subsection{Photometric errors}\label{sec:photerr}

The photometric accuracy  was analysed by computing the
difference between the measured magnitudes and the input ones.  
The differences between the observed and the input magnitude, that
will be referred to as {\it photometric errors}, are strongly dependent on
the crowding conditions and on the magnitude of the star.  When crowding is
negligible, photometric errors are dominated by the statistic of the counts and their distribution
is expected to
be symmetric, with null median value and a dispersion increasing
with the input magnitudes. As the stellar crowding grows,
the probability of blending of two or more sources increases; consequently
the magnitudes of the stars are more frequently recovered  brighter 
than input ones
and the resulting photometric error distribution is skewed 
towards negative values
\citep[see e.g.][]{greg+2011}.  In this section we 
study the photometric error distribution as a function of
the radial distance from the NSC  centre; in particular we
will determine at which distance from the cluster centre crowding conditions prevent
 reliable photometric measurements.

For each simulation, we divided the stars in three groups,
according to the radial distance from the cluster centre.
We computed 
the median of the photometric errors and the
$1 \sigma$ widths of the error distributions 
binning data in 0.2 mag intervals
(excluding bins with less than 50 stars). The dispersions of the errors in each magnitude bin were computed
independently for the stars measured brighter ($\sigma_-$) and those
measured fainter ($\sigma_+$) than the input magnitudes.  On the basis
of the arguments presented above, the median and the $\sigma_-$ values
trace the effect of blending, while $\sigma_+$ is an indicator of the
contribution to the photometric uncertainties of random factors,
i.e. mainly photon and read-out noise.

\myfigure{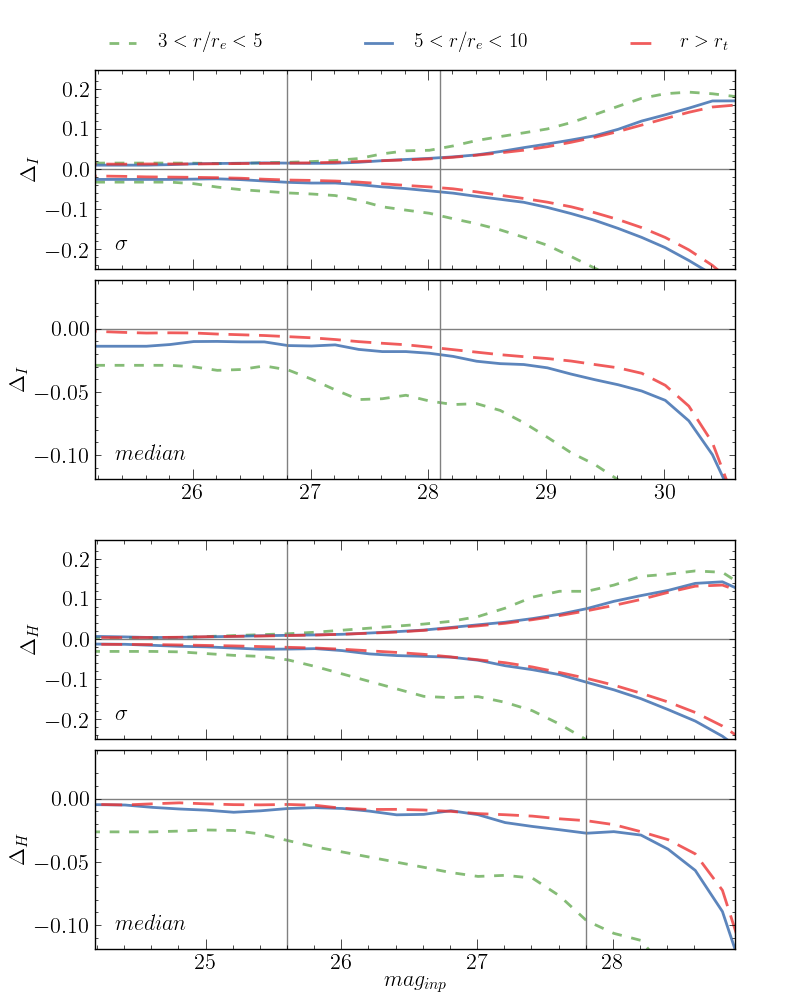} {Photometric errors in the simulation at
  2 Mpc with the 1 Gyr old clusters in $I$ and $H$-band.  The first and
  third panels from the top show the $\sigma_+$ ({\it upper curves})
  and $\sigma_-$ ({\it lower curves}) values for each radial distance
  bin.  The other two panels show the median error.  The two vertical
  lines show the magnitude of the RC and the MSTO.}{fig:err}

Figure \ref{fig:err} shows the photometric errors obtained for the
simulated images with the 1 Gyr old cluster at 2 Mpc.  This case is
used as an example, and the following considerations as well as the
general results, apply also to the other simulations.  We
note that the error distribution in the outer region of the cluster
($5<r/r_e <10$) is almost identical to that in the very external
regions, beyond the cluster tidal radius ($r>r_t$), over the whole
magnitude range probed.  This means that the photometric accuracy in
the outer part of the cluster is the same as in the region beyond the
tidal radius, where no cluster members are present. In other words,
the crowding affects the photometric errors only at radial distances
$r\lesssim 5\,r_e$.  It is interesting to note that photometric errors
for the brightest sources are smaller in the reddest bands; the
$\sigma_+$ and $\sigma_-$ are in fact $\sim 0.03$ mag in the $I$ band,
and $\lesssim 0.01$ mag in $H$.  
This is because the AO correction is more efficient
at longer wavelengths,
leading to a higher Strehl ratio in redder bands
\citep{cili+2012}. Moreover the $I$-band images are
slightly under-sampled (the FWHM of the PSF core in the $I$-band is $\sim 1.8$
pixels).  On the other hand, the photometric errors for faintest stars
--at the level of the MSTO-- are smaller in the blue bands.  This is
due to the lower background noise in the $I$-band images.  

\begin{figure}  
  \resizebox{\hsize}{!}{\includegraphics{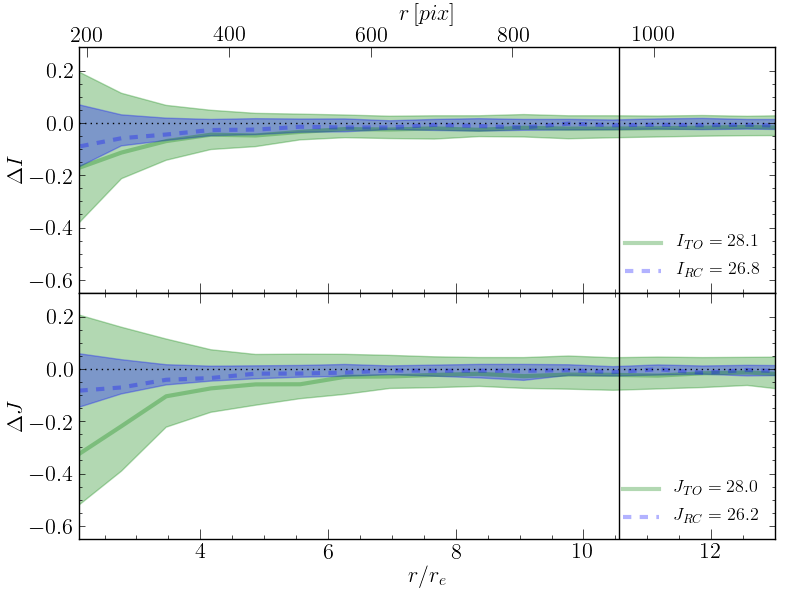}}
  \resizebox{\hsize}{!}{\includegraphics{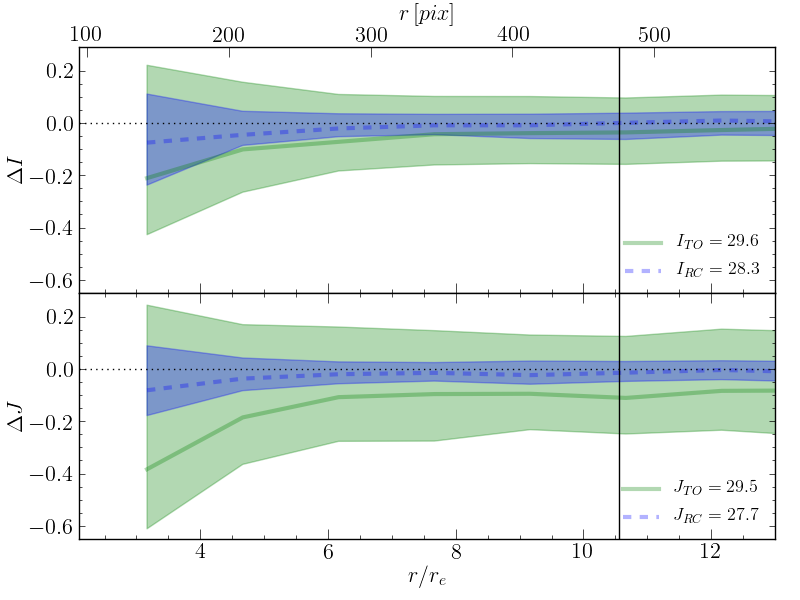}}
  \caption{Photometric errors for the simulation at 2 Mpc ({\it upper panels}) and 4 Mpc ({\it lower panels})
    in the $I$ and $J$ bands for stars at magnitudes corresponding to the 1 Gyr MS turn-off and red-clump. The vertical line corresponds to
  the NSC tidal radius. }
  \label{fig:err2}
\end{figure}

Figure \ref{fig:err2} shows the dependence on distance from the
cluster centre of the width of the error distribution at the MSTO
magnitude and at the clump magnitude for the case of a 1 Gyr old SSP
at 2 and 4 Mpc distance.  As seen before the dispersion of the
photometric errors distribution decreases at increasing radial
distances in the central region of the cluster. In the outer region
the stellar surface density is lower and therefore the effects of
stellar blending on the photometric accuracy are negligible. As a
consequence, beyond a critical radial distance the photometric errors
(both $\sigma_-$ and $\sigma_+$) are constant.  For MSTO stars this
critical radial distance is $\sim 5 r_e$ at 2 Mpc and $\sim 6 r_e$ at
4 Mpc for the 1 Gyr old case. For the 4 Gyr NSC the critical
radius for MSTO stars is larger, because the MSTO is fainter. Specifically the radius is $\sim
6 r_e$ and $\sim7 r_e$ respectively for the simulations at 2 and 4 Mpc. For the 10 Gyr old NSC
the MSTO at 4 Mpc is not detected, while at 2 Mpc it is at the
detection limit.

\subsection{Completeness}

\myfigure{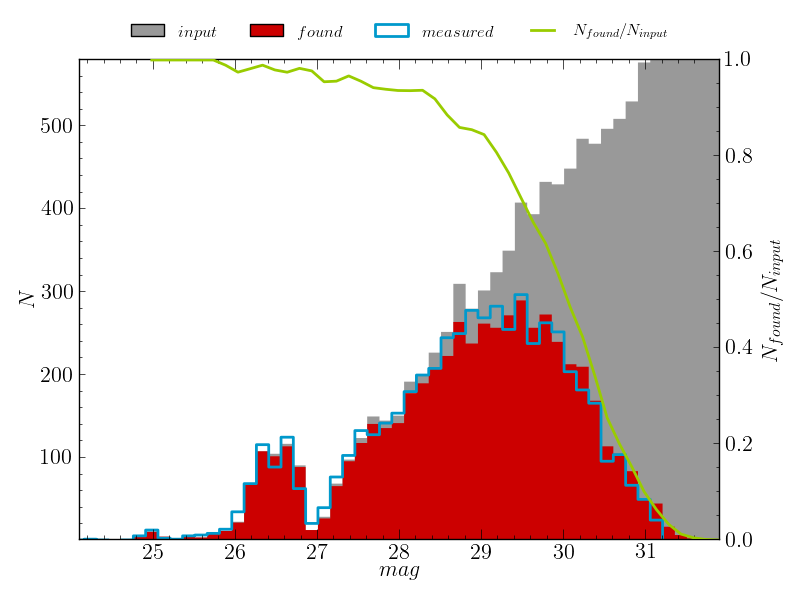}{
\referee{$I$-band} luminosity function of all input stars ({\it grey}) and of input stars
position-matched with a source in the output catalogue ({\it red}).  We
also plotted the LF obtained from the measured magnitudes in the
output catalogue ({\it blue}). The completeness function is shown by the
{\it green} line and its scale is shown 
on the right-hand axis.
The LFs are computed from 
stars with a
radial distance between $3.5 r_e$ and $4.5 r_e$ in the images of
the 1 Gyr old cluster at 2~Mpc.}{fig:lf}

Figure \ref{fig:lf} shows stellar luminosity
functions (LF) for the simulation of the 1 Gyr old cluster at 2
Mpc. Specifically, the {\it input} luminosity function is shown as a
grey histogram; the red histogram is the LF of all input stars that
were {\it found}, i.e. position-matched with a source detected by
Starfinder. This LF is plotted as function of the input magnitudes.  The open
blue histogram is the output LF, function of the {\it measured}
magnitude.  The difference between the last two LFs is due to
photometric errors.  We define as completeness factor the ratio
between the first two LFs, $N_{found}/N_{input}$. We note that 
no selection
on the photometric error was \referee{applied}
to build the LFs used to define the completeness. 
The completeness is therefore the
probability that a star with a given input magnitude is recovered by
our reduction procedure, i.e. matched with a measured star, irrespective of its photometric error.   
The completeness curve is shown as a green
line in Fig. \ref{fig:lf} and its scale is drawn on the right-hand
axis.

\begin{figure}
  \resizebox{\hsize}{!}{\includegraphics{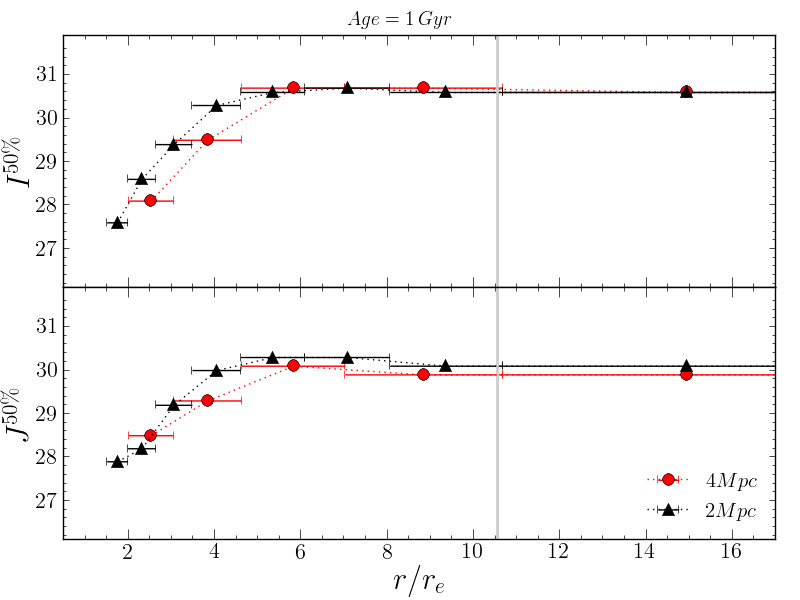}}\\[1em]
  \resizebox{\hsize}{!}{\includegraphics{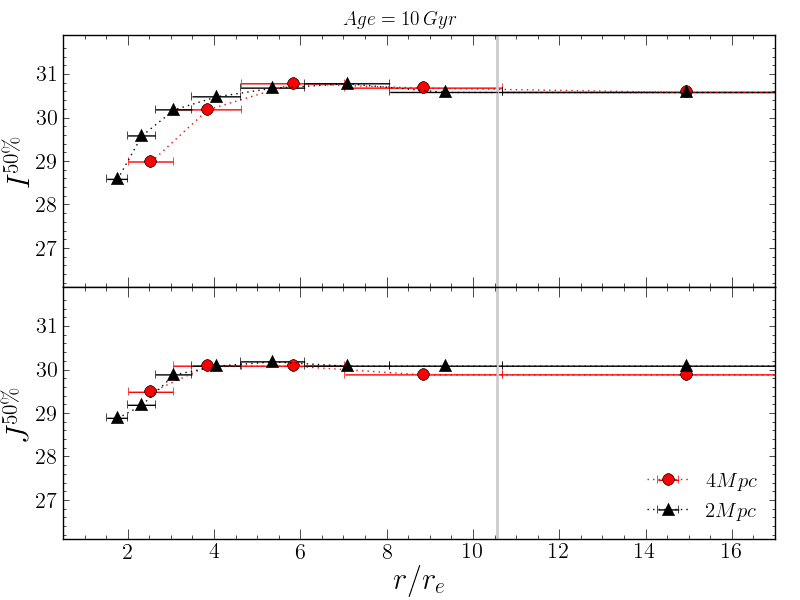}}
  \caption{The 50\% completeness magnitude as
  a function of the radial distance from the NSC centre. The vertical
  line is the NSC tidal radius. Results for the simulations with
  the 1 and 10 Gyr old clusters are shown in the upper and lower panels respectively.}
  \label{fig:compl}
\end{figure}

We characterise the photometric depth as the magnitude level at which
a $50\%$ completeness factor is reached (mag$^{50\%})$.  Figure
\ref{fig:compl} shows mag$^{50\%}$ as a function of the radial
distance from the NSC centre for the simulations with the young star
cluster.  At small radial distances from the cluster centre the
photometry is clearly shallower than in the external regions because
of the extreme crowding conditions.  At increasing radial distances
the stellar density decreases and the photometry is progressively
deeper. Beyond a certain radius the stellar crowding is not the
dominant factor that determine the photometric depth and therefore
mag$^{50\%}$ is constant.  The radial distance at which the maximum
photometric depth is reached depends on the distance and age of the
NSC.  
For the 1 Gyr old NSC it is $\sim5 r_e$ at
2 Mpc and $\sim6 r_e$ at 4 Mpc. 
Comparing the simulations with ages of 1 and 10 Gyr in Fig. \ref{fig:compl}
the completeness is higher for the older SSP. Specifically,
the maximum photometric depth is reached at smaller radii, and
mag$^{50\%}$ at fixed radius is fainter, for the older SSP.
A consistent result applies to the 4 Gyr old SSP, for which we find 
a completeness level intermediate between the 1 and 10 Gyr old
simulations. This trend is likely the consequence of the smaller
number of stars in the images with older NSC (see also Fig. \ref{fig:cmds}),
which reduces the effect of crowding on photometry.

The magnitudes of the MSTO of the 1 Gyr-old NSC at 2 Mpc are $I=28.1$ and
$J=28.0$ at 2 Mpc. The corresponding magnitudes at 4 Mpc are 1.5 mag
fainter. Figure \ref{fig:compl} therefore shows that the photometric
completeness of MSTO stars is above 50\% at radial distance
$\gtrsim2r_e$ in the simulations of stellar systems at 2 Mpc and
$\gtrsim3r_e$ in the simulations at 4 Mpc.

The mag$^{50\%}$ value at fixed $r/r_e$ is fainter in the simulations
at 2 Mpc compared to those at 4 Mpc (see Fig. \ref{fig:compl}). 
This reflects the worse crowding 
conditions, due  to the higher surface density, which affects the simulations
at larger distances.

\subsection{Stellar populations}\label{sec:stelpops}

\begin{figure*}    
\centering
  \includegraphics[width=3.5cm]{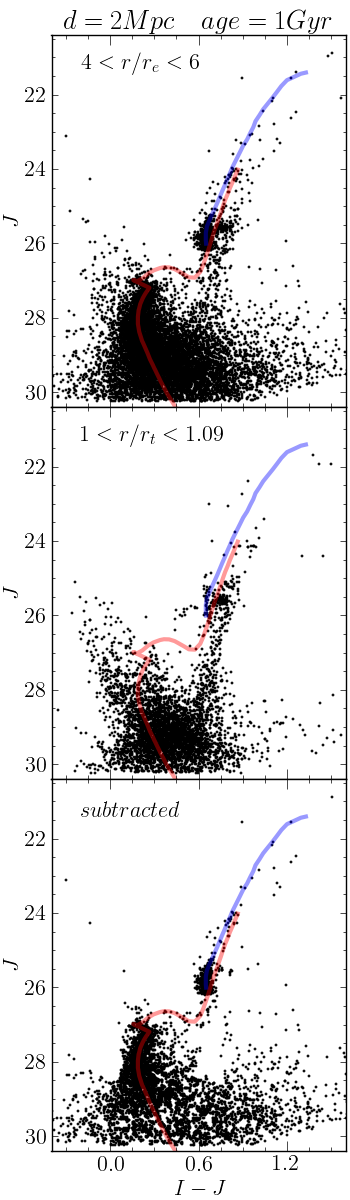}\hfill
  \includegraphics[width=3.5cm]{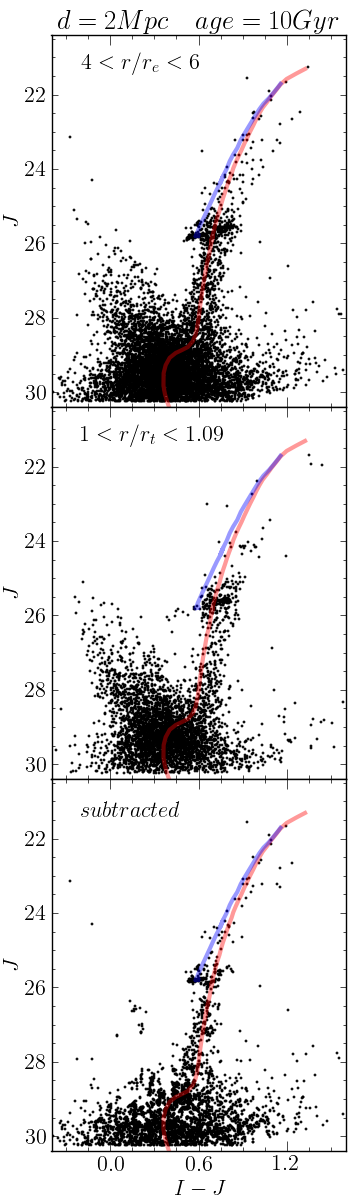}\hfill
  \includegraphics[width=3.5cm]{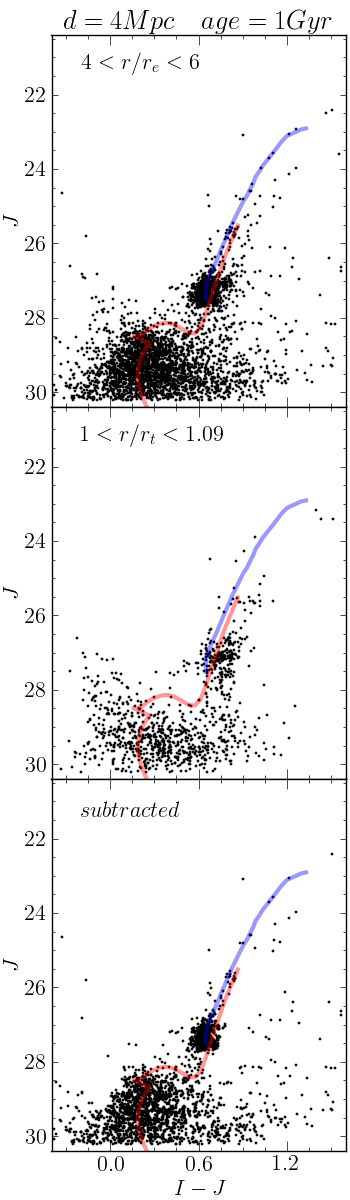}\hfill
  \includegraphics[width=3.5cm]{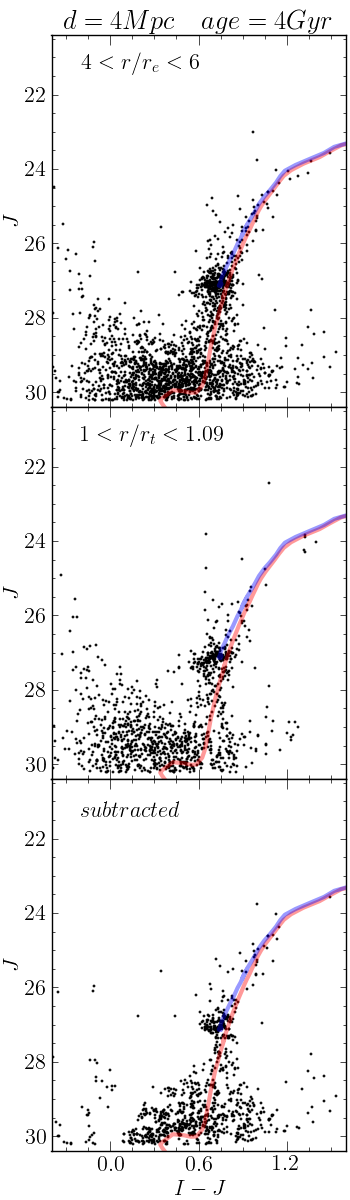}\hfill
  \includegraphics[width=3.5cm]{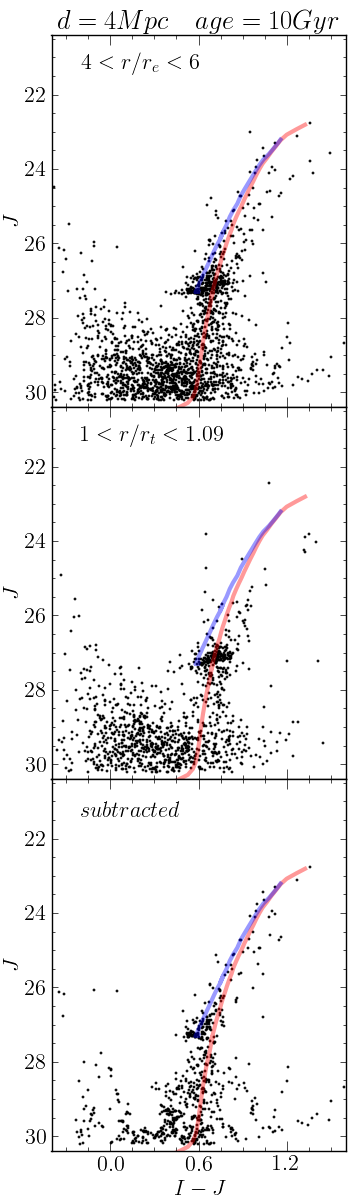}
  \caption{Statistical subtraction of host galaxy stellar members from the CMDs of NSC stellar populations.
    Only few representative cases are shown.
    The {\it upper panels} show the CMDs in the target region. The CMDs of control field beyond the NSC tidal radius are shown in {\it central panels}.
    The resulting decontaminated CMDs are shown in the {\it lower panels}.
    In each panel we show stellar models from \cite{mari+2008}.
    The {\it red line} is the MS and the RGB, while the {\it blue line}
    is the sequence of core He-burning stars, up to the first
    AGB thermal pulse.}
  \label{fig:subcmd}
\end{figure*}

In this section we  analyse in more details the CMDs and in
particular we discuss whether they can be used for a reliable
measurement of the age of the NSC.  The photometry in the $H$ and
$K$-band images is shallower than in the bluer bands \citep[see
  also][]{greg+2012,schr+2014}; we derived mag$^{50\%} \sim27.5$ mag in the $K$ band,
  corresponding to the MSTO of the 1 Gyr old NSC at 2 Mpc.
  In all other cases we did not obtained reliable
magnitude measures for MSTO stars in these bands. In this section we therefore
consider only $I$ and $J$-band photometry.

As mentioned before, the NSCs are embedded in the dense central regions
of the host galaxy and therefore NSC stellar populations must be disentangled from the 
host galaxy members. To assess this issue we
performed a statistical subtraction of the contaminating host galaxy
stellar populations.  We excluded from our analysis the central
regions of the clusters, where the photometry is not reliable because
of stellar blends. We also excluded the very external regions of the
clusters, where the number of member stars is much lower than the
number of non-members.  
We performed some tests to determine   the optimal region and found that the best results
are achieved considering the ring
between 4 and 6 $r_e$.
As a
control field, populated by host galaxy members only, we used the
region located between 1 and 1.09 $r_t$. This region has an area equal to the
area of the target region 
(we remind that $r_t/r_e=10.56$).  We therefore assumed that the stellar population in this control field is the same as
 the stellar population of the host galaxy in the target field.
 For each star
in the control field, we subtracted a star in the target field.  The
star to be subtracted is chosen as the closest in the CMD diagram; following
\cite{zocc+2003} this was selected as the star with the minimum value
of
\begin{equation}
d=\sqrt{
[5 \Delta(I-J)]^2+\Delta J^2
}.
\end{equation}

Fig. \ref{fig:subcmd} shows the total, control-field, and field-subtracted CMDs.
The isochrones 
used to simulate the
NSC stellar populations are shown to highlight the main evolutionary sequences.
The cluster stellar populations is clearly visible in all the five CMDs shown 
in the bottom panels of Fig. \ref{fig:subcmd}, in which the detection of the NSC stellar population is much more reliable than in the original CMDs.
In particular, we notice the case of the 10 Gyr old cluster at 4 Mpc: while the top and middle panel of Fig \ref{fig:subcmd}  show very similar CMDs, the result of the
statistical subtraction shows a residual clump and RGB in the bottom
panel, clear sign of the presence of an old stellar population.

On the basis of the CMDs in Fig. \ref{fig:subcmd} we classified the simulated stellar systems into three groups.

\begin{itemize}
\item Group A, in which the detection of the MSTO is clear and the age of the stellar population can be safely obtained.
This is the case for the 1 Gyr cluster located both at 2 and 4 Mpc and for the 4 Gyr cluster at 2 Mpc (not shown in Fig. \ref{fig:subcmd}).

\item Group B, in which the MSTO is at the detection limit and it is therefore not possible to precisely
  estimate the age of the stellar population.
  This is the case for the 10
  Gyr NSC at 2 Mpc and the 4 Gyr NSC at 4 Mpc. The MSTO is at the
  detection limit, at about $J\simeq29$ and 30 mag, respectively. At
  this faint magnitudes 
  photometric errors and systematic biases due to stellar blends
  put strong limitations in the measurement of the MSTO
  magnitude and colour. 
  Nevertheless, it is still possible to derive some approximate information on the age of the stellar population.
  
\item Group C in which, like in the 10 Gyr old  NSC at 4 Mpc, we can
  detect the bright part of the stellar population of the
  NSC. (i,.e. the He burning clump and the RGB stars).  The absence of
  bright blue MS stars is a strong evidence that no young stellar
  populations are present. The colour and extension of the RGB and the
  morphology of the RC can be used to characterise the age of the stellar population.
  To be conservative we can say that in these cases it
  would be possible to obtain just a reliable lower limit for the age
  of the NSC stellar populations.
\end{itemize}

\section{Comparison with JWST and TMT}\label{sec:instruments}

\begin{figure}
  \centering
  \includegraphics[width=7.3cm]{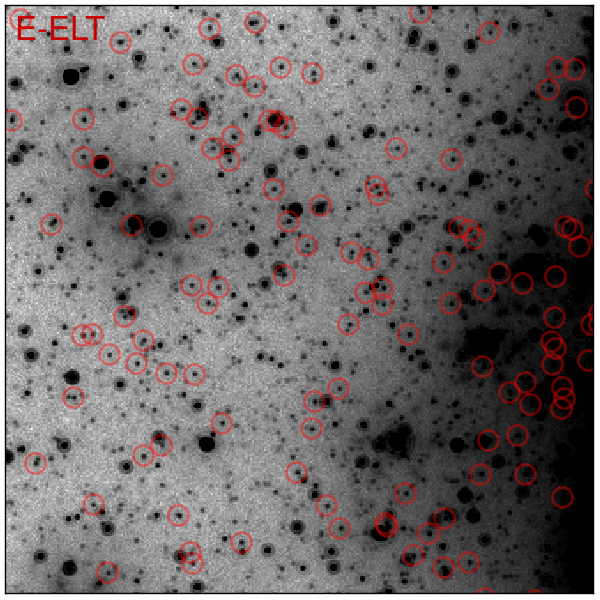}\\
  \includegraphics[width=7.3cm]{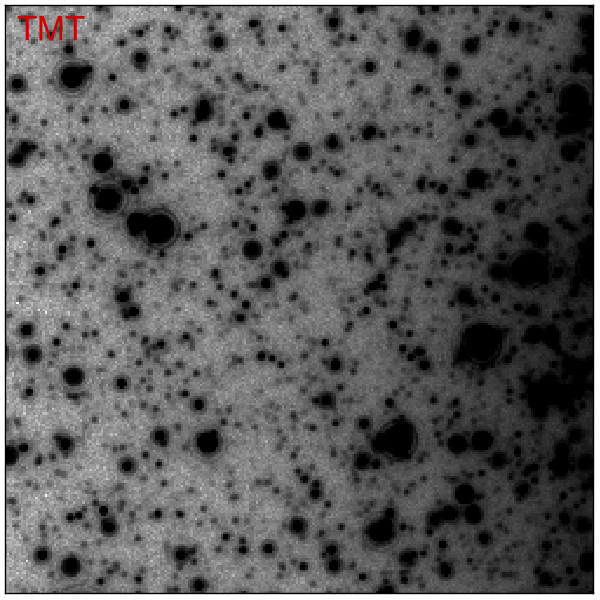}\\
  \includegraphics[width=7.3cm]{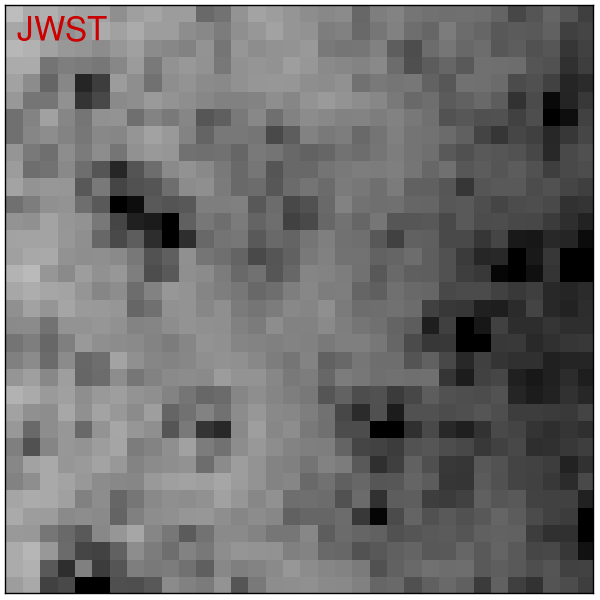}
  \caption{$1\arcsec \times 1\arcsec$ E-ELT ({\it top panel}), TMT
    ({\it middle panel}), and JWST ({\it bottom panel}) images at 6
    $r_e$ from the centre of the 1 Gyr old NSC at 2~Mpc.
    In the top panel the {\it red circles} shows $J=28$ mag stars, corresponding
  to the MSTO magnitude.}
  \label{fig:confima}
\end{figure}

\subsection{JWST}
In this section we presents results obtained using the NIRCam Exposure
Time Calculator Version P1.6\footnote{\url{http://jwstetc.stsci.edu/etc/input/nircam/imaging}}.
The NIRCam limiting Vega-magnitude for an exposure time of $3~h$
results to be $J \sim 28.5$ mag for a $S/N=5$ and $J\sim 27.8$ mag for
$S/N=10$.  For comparison, using the AETC, we obtained that the
corresponding magnitudes for MICADO are $J=29.7$ mag ($S/N=5$) and
$J=28.8$ mag ($S/N$=10).
The MSTO of sufficiently 
young and nearby NSCs could still be detected with JWST, but
crowding would highly hamper the determination of their age. Indeed, the pixel size of NIRCam is
31.7 mas, $\sim 10$ time larger than that of MICADO. Given the
high-spatial stellar densities in NSC, the NIRCam spatial resolution
is therefore not sufficient to obtain accurate photometry of the NSC individual
stars. To illustrate this point,
in the bottom panel of
Fig. \ref{fig:confima} we show a simulated JWST
image, obtained using the NIRCam
PSF\footnote{\url{http://www.stsci.edu/~mperrin/software/psf_library/}}.

\referee{ Spectral info for the brightest stars of the cluster will
  probably be achievable with the JWST near-infrared multi-object
  spectrograph NIRSpec. This could be useful to furthermore characterise the
  properties of the NSC stellar populations.  }
\subsection{IRIS@TMT}

An example of a $J$-band IRIS simulated image is shown in
Fig. \ref{fig:confima}.  This image does not show 
significant difference with respect to the one obtained
from MICADO simulations. we used Starfinder
photometry obtained from the simulated image
to derive a quantitative comparison between
the performances of the IRIS and MICADO.
The filter set of the two instruments
is not the same, in particular IRIS will not be equipped with a
$I$-band filter but a $Z$-band filter will be mounted. 
As the effective wavelength of $I$ ($\lambda=0.90\, \mu m$) and $Z$ ($\lambda=0.93\, \mu m$) bands are similar,
we  compare IRIS $Z$-band photometry with MICADO $I$-band one.

\begin{figure}
  \centering
      \resizebox{\hsize}{!}{\includegraphics{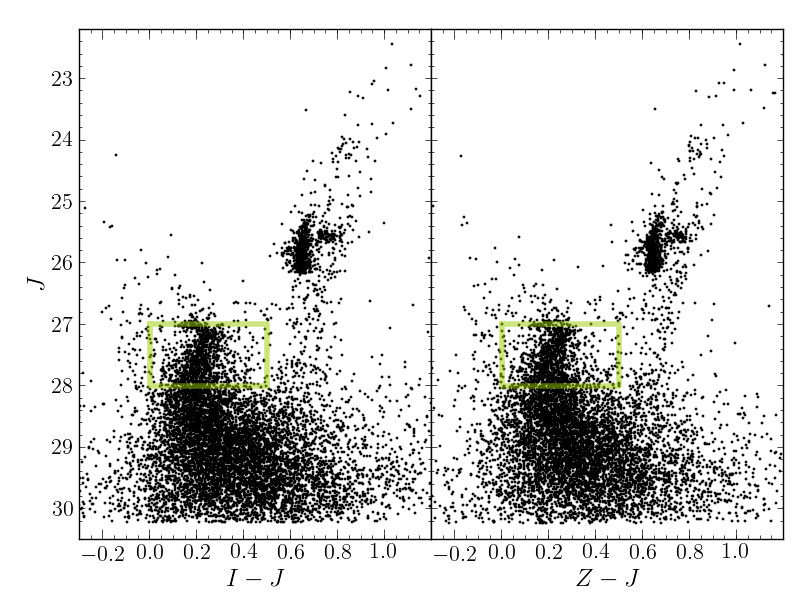}}\\
      \resizebox{\hsize}{!}{\includegraphics{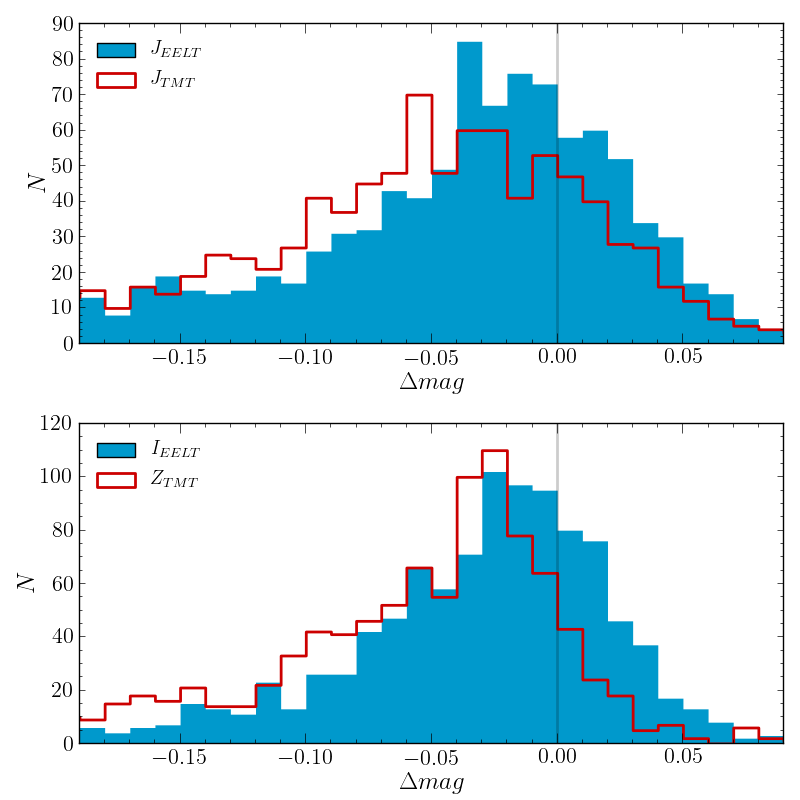}}
  \caption{{\it Top panels:} CMDs for stars in the $4<r/r_e<6$ region of \mbox{MICADO} ({\it
      left panel}) and IRIS ({\it right panel}) simulated images.
The green rectangle shows the region selected for the comparison of the photometric errors shown in the panels below.
   {\it Middle Panel:} photometric errors 
  for MSTO stars in  $4<r/r_e<6$ region of $J$-band MICADO and IRIS image.
  {\it Lower panel:} the same for MICADO $J$ and IRIS $Z$-band images.
  MSTO have been selected in the box shown in the top panels. 
}
  \label{fig:compcmd}
\end{figure}

Figure \ref{fig:compcmd} compares the CMDs obtained from the IRIS and
\mbox{MICADO} simulated images. The magnitude limits and broadening of
the evolutionary sequences in the two CMDs indicate that the
photometric depth and accuracy of the two sets are comparable.  The
comparison of IRIS and MICADO photometric errors for MSTO stars is
shown in the middle and lower panel of Fig. \ref{fig:compcmd}.  The
distributions of the differences between measured and input magnitudes
are very similar for TMT and E-ELT photometry in both $J$ and $I$ (or
$Z$) band.  The only difference is that the TMT's distributions are
shifted towards negative values.  This effect is very small
($0.01--0.02$ mags) and is likely related to the higher contribution
of stellar blends in the TMT photometric errors. The FWHM of MICADO
PSF is in fact 5.1 mas in $I$-band and 6.3 mas in the $J$-band, while
the FWHM of IRIS PSFs are 8.8 mas and 9.5 mas in $Z$ and $J$-band
respectively.

\begin{figure}
  \resizebox{\hsize}{!}{\includegraphics{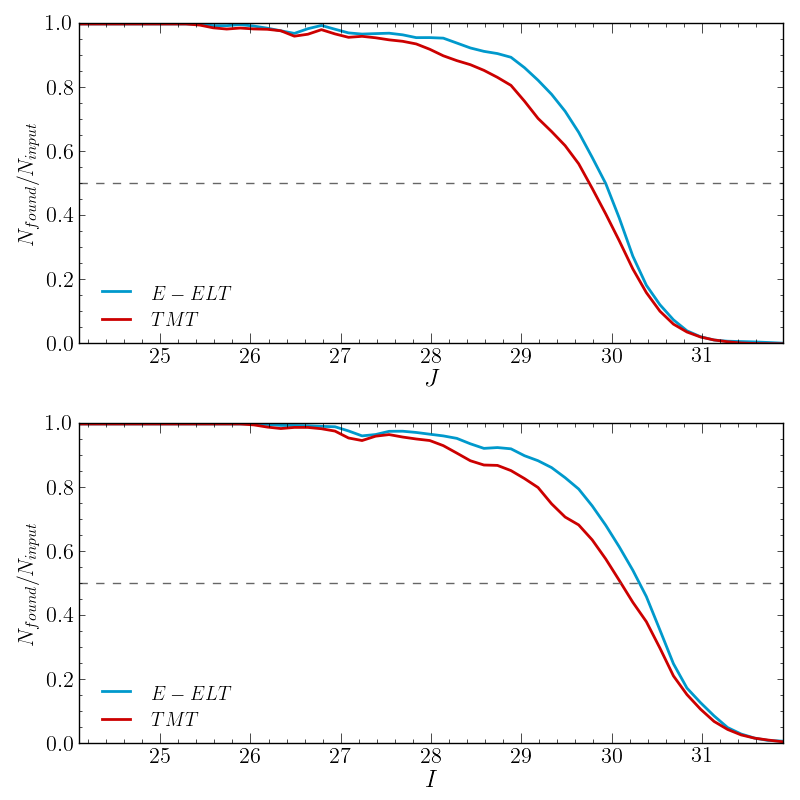}}
  \caption{Completeness of E-ELT and TMT $I$-band photometry.}
  \label{fig:compcompl}
\end{figure}

The photometric completeness of IRIS photometry is compared to
the one of MICADO in Fig. \ref{fig:compcompl}. The two completeness curves
 are very similar. The
$J$-band 50\% completeness in MICADO photometry is found at magnitude
$\sim 0.2$ mag fainter than in IRIS photometry. The same is found
when comparing MICADO $I$-band with IRIS $Z$-band photometry.

On the one hand the E-ELT collecting area is larger than the
one  of TMT; on the other hand the MAORY PSF Strehl ratio is somewhat
lower than the one foreseen for NFIRAOS \citep{cili+2012,do+2014}.  The two
effects compensate each other and the performances of E-ELT and TMT
imagers result to be very similar for the particular science
application considered in this paper.

\section{Summary and conclusions}\label{sec:conclusions}

In this paper we present simulated observations of NSCs in nearby
galaxies.  The simulated images were analysed using the Starfinder
photometric package and the resulting CMDs were studied to assess the
feasibility of the accurate study of the NSC stellar populations.  
For simplicity, we modelled NSCs with SSPs, but our results are applicable
also \referee{to} NSC originated from extended star-formation episodes; these complex stellar populations
can be considered as the superposition of multiple SSPs.
We constructed synthetic images adopting the parameters
currently foreseen for the MICADO camera at E-ELT, exploring a number
of cases with different age and distance of the NSC. We then proceeded
with the photometric measurements of the images, the construction of
the CMDs, and their analysis.

Our results are summarised in Fig. \ref{fig:maglim}.  We showed that
E-ELT photometry could be used to age-date NSCs up to 10 Gyr in
galaxies at 2 Mpc. Note, however, that the MSTO for the oldest stellar
populations falls close to the detection limit and therefore the age
estimates for the oldest stars will be uncertain.  
We \referee{evaluate} these uncertainties 
using isochrones \citep{mari+2008} to derive a relation between cluster age
and MSTO magnitude. For the 10 Gyr old cluster at 2 Mpc
the photometric error is $\sigma_J=0.29$ mag for MSTO stars, corresponding to
an age uncertainty $\sigma [\log(t)]=0.10$,
--i.e. $t=10.0^{+2.6}_{-2.1}$ Gyr. For galaxies at 4 Mpc the age of the
NSC stellar population can be estimated with high accuracy only for
young stellar populations.
For oldest stellar populations the MSTO is detectable up to 4 Gyr.
\referee{At this distance the MSTO of 4 Gyr-old stellar populations
is detectable with}
 a photometric uncertainty $\sigma_J=0.35$ mag,
corresponding to $\sigma [\log(t)]=0.12$ --i.e. $t=4.0^{+1.3}_{-1.0}$
Gyr.  

We can now extrapolate our results to
galaxies located at different distances.  In this paper we showed that
E-ELT imaging can provide stellar photometry complete at a 50\% level at
$J\simeq30.0$ mag. For stars 0.5 mag brighter than this the completeness reaches
$\sim 80\%$ level.  Using the relation between cluster age and MSTO magnitude,
 we
calculated 
the age of SSPs whose MSTO is at $J=30.0$ mag (and $J=29.5$ mag) as a function of distance modulus.
These relations are shown in Fig. \ref{fig:maglim} and we use them 
to extrapolate the results of our simulations.
Stellar populations in the green area --below the $J=28.5$mag curve-- will have a
MSTO detectable above the $\sim 80\%$ completeness limit; for all these
stellar populations the age will be therefore determined with a high
reliability. 
In the case of stellar populations in the orange area,
the photometric completeness for MSTO
stars is estimated to be 50\%-80\%; 
in this case age measurements 
are expected to be affected by large uncertainties.

\myfigure{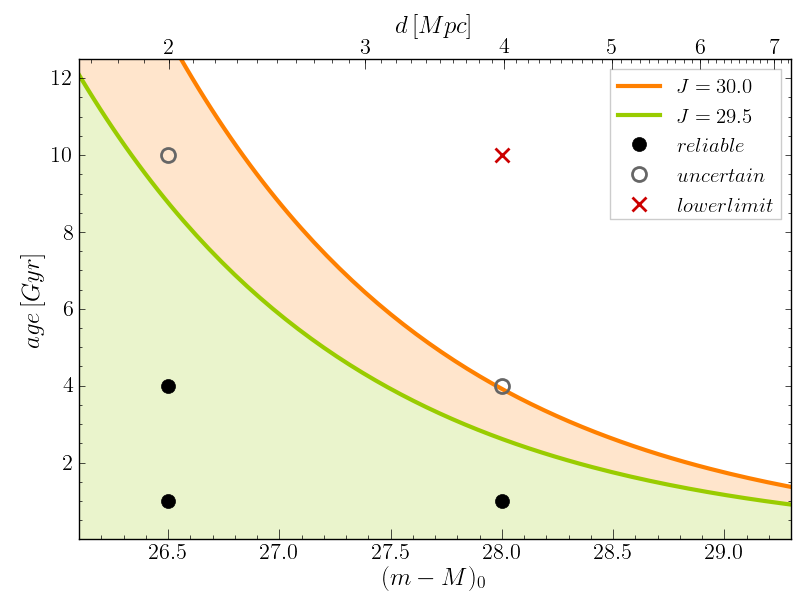}{
Age of the oldest MSTO detectable as a function of the target distance.
We show the results obtained adoption a detection magnitude limit 
of 30.0 and 29.5 in the $J$-band. These are the magnitudes corresponding to the 50\% and 80\% photometric completeness level.
}{fig:maglim}

We finally investigated the possibility to obtain reliable photometry
for resolved stellar populations in NSC with other next-generation
instrumentation.  We showed that the spatial resolution of the JWST is
not sufficient to resolve nearby NSC into individual
stars. Simulations of TMT images show that for our specific science
case TMT and E-ELT performances are nearly equivalent and therefore
the results of our analysis with the E-ELT can be applied also to TMT
observations. To conclude, we showed that future ground-based 30 meter
class telescopes, equipped with advanced Adaptive Optics systems, will
provide accurate photometric measurements of the resolved stars that
will allow us to directly probe the formation and evolution of the
stellar populations in NSC.

In the Cosmicflows-2 compilation of nearby galaxies with measured
distance \citep{tull+2013} there are 113 galaxies with distance
between 2 and 4 Mpc.  
The estimate of the number of galaxies hosting a NSC is $\sim 70\%$
\citep{cote+2006,turn+2012}.  
It turns out therefore that the future facilities like E-ELT and TMT
--that together are able to cover the whole sky-- will provide the
fundamental data to unveil the mechanisms of formation of the nuclear
clusters of galaxies.

\begin{acknowledgements}
We acknowledge  support of INAF and MIUR through the grant {\it Progetto premiale T-REX}.
We warmly thank E. Diolaiti for the useful discussions. We thank S. Wright, T. Do, L. Wang and B. Ellerbroek
for sharing their IRIS PSFs.
\end{acknowledgements}
\bibliographystyle{aa}
\bibliography{eelt_clusters}

\begin{thebibliography}{34}
\expandafter\ifx\csname natexlab\endcsname\relax\def\natexlab#1{#1}\fi

\bibitem[{{B{\"o}ker}(2010)}]{boke+2010}
{B{\"o}ker}, T. 2010, in IAU Symposium, Vol. 266, IAU Symposium, ed. R.~{de
  Grijs} \& J.~R.~D. {L{\'e}pine}, 58--63

\bibitem[{{B{\"o}ker} {et~al.}(2002){B{\"o}ker}, {Laine}, {van der Marel},
  {Sarzi}, {Rix}, {Ho}, \& {Shields}}]{boke+2002}
{B{\"o}ker}, T., {Laine}, S., {van der Marel}, R.~P., {et~al.} 2002, \aj, 123,
  1389

\bibitem[{{B{\"o}ker} {et~al.}(2004){B{\"o}ker}, {Sarzi}, {McLaughlin}, {van
  der Marel}, {Rix}, {Ho}, \& {Shields}}]{boke+2004}
{B{\"o}ker}, T., {Sarzi}, M., {McLaughlin}, D.~E., {et~al.} 2004, \aj, 127, 105

\bibitem[{{Capuzzo-Dolcetta} \& {Miocchi}(2008)}]{capu+2008}
{Capuzzo-Dolcetta}, R. \& {Miocchi}, P. 2008, \apj, 681, 1136

\bibitem[{{Ciliegi} {et~al.}(2012){Ciliegi}, {Diolaiti}, {Baruffolo},
  {Bellazzini}, {Bregoli}, {Butler}, {Conan}, {Cosentino}, {Foppiani},
  {Lombini}, {Marchetti}, {Petit}, {Robert}, {Schreiber}, {Biliotti},
  {D'Odorico}, {Fusco}, \& {Hubin}}]{cili+2012}
{Ciliegi}, P., {Diolaiti}, E., {Baruffolo}, A., {et~al.} 2012, \memsai, 83,
  1151

\bibitem[{{C{\^o}t{\'e}} {et~al.}(2006){C{\^o}t{\'e}}, {Piatek}, {Ferrarese},
  {Jord{\'a}n}, {Merritt}, {Peng}, {Ha{\c s}egan}, {Blakeslee}, {Mei}, {West},
  {Milosavljevi{\'c}}, \& {Tonry}}]{cote+2006}
{C{\^o}t{\'e}}, P., {Piatek}, S., {Ferrarese}, L., {et~al.} 2006, \apjs, 165,
  57

\bibitem[{{Davies} {et~al.}(2010){Davies}, {Ageorges}, {Barl}, {Bedin},
  {Bender}, {Bernardi}, {Chapron}, {Clenet}, {Deep}, {Deul}, {Drost},
  {Eisenhauer}, {Falomo}, {Fiorentino}, {F{\"o}rster Schreiber}, {Gendron},
  {Genzel}, {Gratadour}, {Greggio}, {Grupp}, {Held}, {Herbst}, {Hess},
  {Hubert}, {Jahnke}, {Kuijken}, {Lutz}, {Magrin}, {Muschielok}, {Navarro},
  {Noyola}, {Paumard}, {Piotto}, {Ragazzoni}, {Renzini}, {Rousset}, {Rix},
  {Saglia}, {Tacconi}, {Thiel}, {Tolstoy}, {Trippe}, {Tromp}, {Valentijn},
  {Verdoes Kleijn}, \& {Wegner}}]{davi+2010}
{Davies}, R., {Ageorges}, N., {Barl}, L., {et~al.} 2010, in Proc. SPIE, Vol.
  7735, Ground-based and Airborne Instrumentation for Astronomy III, ed. I.~S.
  {McLean}, S.~K. {Ramsay}, \& H.~{Takami}, id. 77352A

\bibitem[{{Deep} {et~al.}(2011){Deep}, {Fiorentino}, {Tolstoy}, {Diolaiti},
  {Bellazzini}, {Ciliegi}, {Davies}, \& {Conan}}]{deep+2011}
{Deep}, A., {Fiorentino}, G., {Tolstoy}, E., {et~al.} 2011, \aap, 531, A151

\bibitem[{{Diolaiti} {et~al.}(2000){Diolaiti}, {Bendinelli}, {Bonaccini},
  {Close}, {Currie}, \& {Parmeggiani}}]{diol+2000}
{Diolaiti}, E., {Bendinelli}, O., {Bonaccini}, D., {et~al.} 2000, in Proc.
  SPIE, Vol. 4007, Adaptive Optical Systems Technology, ed. P.~L. {Wizinowich},
  879

\bibitem[{{Diolaiti} {et~al.}(2010){Diolaiti}, {Conan}, {Foppiani},
  {Marchetti}, {Baruffolo}, {Bellazzini}, {Bregoli}, {Butler}, {Ciliegi},
  {Cosentino}, {Delabre}, {Lombini}, {Petit}, {Robert}, {Rossettini},
  {Schreiber}, {Tomelleri}, {Biliotti}, {D'Odorico}, {Fusco}, {Hubin}, \&
  {Meimon}}]{diol+2010}
{Diolaiti}, E., {Conan}, J.-M., {Foppiani}, I., {et~al.} 2010, in Proc. SPIE,
  Vol. 7736, Adaptive Optics Systems II, ed. B.~L. {Ellerbroek}, M.~{Hart},
  N.~{Hubin}, \& P.~L. {Wizinowich}, id. 77360R

\bibitem[{{Do} {et~al.}(2014){Do}, {Wright}, {Barth}, {Barton}, {Simard},
  {Larkin}, {Moore}, {Wang}, \& {Ellerbroek}}]{do+2014}
{Do}, T., {Wright}, S.~A., {Barth}, A.~J., {et~al.} 2014, \aj, 147, 93

\bibitem[{{Falomo} {et~al.}(2011){Falomo}, {Fantinel}, \&
  {Uslenghi}}]{falo+2011}
{Falomo}, R., {Fantinel}, D., \& {Uslenghi}, M. 2011, in Proc. SPIE, Vol. 8135,
  Applications of Digital Image Processing XXXIV, ed. A.~G. {Tescher}, id.
  813523

\bibitem[{{Ferrarese} {et~al.}(2006){Ferrarese}, {C{\^o}t{\'e}}, {Dalla
  Bont{\`a}}, {Peng}, {Merritt}, {Jord{\'a}n}, {Blakeslee}, {Ha{\c s}egan},
  {Mei}, {Piatek}, {Tonry}, \& {West}}]{ferr+2006}
{Ferrarese}, L., {C{\^o}t{\'e}}, P., {Dalla Bont{\`a}}, E., {et~al.} 2006,
  \apjl, 644, L21

\bibitem[{{Georgiev} \& {B{\"o}ker}(2014)}]{geor+2014}
{Georgiev}, I.~Y. \& {B{\"o}ker}, T. 2014, MNRAS, in press. ArXiv e-prints

\bibitem[{{Greene} {et~al.}(2010){Greene}, {Beichman}, {Gully-Santiago},
  {Jaffe}, {Kelly}, {Krist}, {Rieke}, \& {Smith}}]{gree+2010}
{Greene}, T., {Beichman}, C., {Gully-Santiago}, M., {et~al.} 2010, in Proc.
  SPIE, Vol. 7731, Space Telescopes and Instrumentation 2010: Optical,
  Infrared, and Millimeter Wave, id. 77310C

\bibitem[{{Greggio} {et~al.}(2012){Greggio}, {Falomo}, {Zaggia}, {Fantinel}, \&
  {Uslenghi}}]{greg+2012}
{Greggio}, L., {Falomo}, R., {Zaggia}, S., {Fantinel}, D., \& {Uslenghi}, M.
  2012, \pasp, 124, 653

\bibitem[{{Greggio} \& {Renzini}(2011)}]{greg+2011}
{Greggio}, L. \& {Renzini}, A. 2011, {Stellar Populations. A User Guide from
  Low to High Redshift} (Wiley-VCH Verlag)

\bibitem[{{Larkin} {et~al.}(2010){Larkin}, {Moore}, {Barton}, {Bauman}, {Bui},
  {Canfield}, {Crampton}, {Delacroix}, {Fletcher}, {Hale}, {Loop}, {Niehaus},
  {Phillips}, {Reshetov}, {Simard}, {Smith}, {Suzuki}, {Usuda}, \&
  {Wright}}]{lark+2010}
{Larkin}, J.~E., {Moore}, A.~M., {Barton}, E.~J., {et~al.} 2010, in Proc. SPIE,
  Vol. 7735, Ground-based and Airborne Instrumentation for Astronomy III, ed.
  I.~S. {McLean}, S.~K. {Ramsay}, \& H.~{Takami}, id. 773529

\bibitem[{{Marigo} {et~al.}(2008){Marigo}, {Girardi}, {Bressan}, {Groenewegen},
  {Silva}, \& {Granato}}]{mari+2008}
{Marigo}, P., {Girardi}, L., {Bressan}, A., {et~al.} 2008, \aap, 482, 883

\bibitem[{{Neumayer} \& {Walcher}(2012)}]{neum+2012}
{Neumayer}, N. \& {Walcher}, C.~J. 2012, Advances in Astronomy, 2012

\bibitem[{{Paudel} {et~al.}(2011){Paudel}, {Lisker}, \&
  {Kuntschner}}]{paud+2011}
{Paudel}, S., {Lisker}, T., \& {Kuntschner}, H. 2011, \mnras, 413, 1764

\bibitem[{{Rossa} {et~al.}(2006){Rossa}, {van der Marel}, {B{\"o}ker},
  {Gerssen}, {Ho}, {Rix}, {Shields}, \& {Walcher}}]{ross+2006}
{Rossa}, J., {van der Marel}, R.~P., {B{\"o}ker}, T., {et~al.} 2006, \aj, 132,
  1074

\bibitem[{{Schreiber} {et~al.}(2012){Schreiber}, {Diolaiti}, {Sollima},
  {Arcidiacono}, {Bellazzini}, {Ciliegi}, {Falomo}, {Foppiani}, {Greggio},
  {Lanzoni}, {Lombini}, {Montegriffo}, {Dalessandro}, \& {Massari}}]{schr+2012}
{Schreiber}, L., {Diolaiti}, E., {Sollima}, A., {et~al.} 2012, in Proc. SPIE,
  Vol. 8447, Adaptive Optics Systems III, ed. B.~L. {Ellerbroek},
  E.~{Marchetti}, \& J.~P. {Véran}, id. 84475V

\bibitem[{{Schreiber} {et~al.}(2014){Schreiber}, {Greggio}, {Falomo},
  {Fantinel}, \& {Uslenghi}}]{schr+2014}
{Schreiber}, L., {Greggio}, L., {Falomo}, R., {Fantinel}, D., \& {Uslenghi}, M.
  2014, \mnras, 437, 2966

\bibitem[{{Scott} \& {Graham}(2013)}]{scot+2013}
{Scott}, N. \& {Graham}, A.~W. 2013, \apj, 763, 76

\bibitem[{{Seth} {et~al.}(2010){Seth}, {Cappellari}, {Neumayer}, {Caldwell},
  {Bastian}, {Olsen}, {Blum}, {Debattista}, {McDermid}, {Puzia}, \&
  {Stephens}}]{seth+2010}
{Seth}, A.~C., {Cappellari}, M., {Neumayer}, N., {et~al.} 2010, \apj, 714, 713

\bibitem[{{Seth} {et~al.}(2006){Seth}, {Dalcanton}, {Hodge}, \&
  {Debattista}}]{seth+2006}
{Seth}, A.~C., {Dalcanton}, J.~J., {Hodge}, P.~W., \& {Debattista}, V.~P. 2006,
  \aj, 132, 2539

\bibitem[{{Tremaine} {et~al.}(1975){Tremaine}, {Ostriker}, \&
  {Spitzer}}]{trem+1975}
{Tremaine}, S.~D., {Ostriker}, J.~P., \& {Spitzer}, Jr., L. 1975, \apj, 196,
  407

\bibitem[{{Tully} {et~al.}(2013){Tully}, {Courtois}, {Dolphin}, {Fisher},
  {H{\'e}raudeau}, {Jacobs}, {Karachentsev}, {Makarov}, {Makarova},
  {Mitronova}, {Rizzi}, {Shaya}, {Sorce}, \& {Wu}}]{tull+2013}
{Tully}, R.~B., {Courtois}, H.~M., {Dolphin}, A.~E., {et~al.} 2013, \aj, 146,
  86

\bibitem[{{Turner} {et~al.}(2012){Turner}, {C{\^o}t{\'e}}, {Ferrarese},
  {Jord{\'a}n}, {Blakeslee}, {Mei}, {Peng}, \& {West}}]{turn+2012}
{Turner}, M.~L., {C{\^o}t{\'e}}, P., {Ferrarese}, L., {et~al.} 2012, \apjs,
  203, 5

\bibitem[{{Walcher} {et~al.}(2006){Walcher}, {B{\"o}ker}, {Charlot}, {Ho},
  {Rix}, {Rossa}, {Shields}, \& {van der Marel}}]{walc+2006}
{Walcher}, C.~J., {B{\"o}ker}, T., {Charlot}, S., {et~al.} 2006, \apj, 649, 692

\bibitem[{{Wehner} \& {Harris}(2006)}]{wehn+2006}
{Wehner}, E.~H. \& {Harris}, W.~E. 2006, \apjl, 644, L17

\bibitem[{{Wright} {et~al.}(2010){Wright}, {Barton}, {Larkin}, {Moore},
  {Crampton}, \& {Simard}}]{wrig+2010}
{Wright}, S.~A., {Barton}, E.~J., {Larkin}, J.~E., {et~al.} 2010, in Proc.
  SPIE, Vol. 7735, Ground-based and Airborne Instrumentation for Astronomy III,
  ed. I.~S. {McLean}, S.~K. {Ramsay}, \& H.~{Takami}, id. 77357P

\bibitem[{{Zoccali} {et~al.}(2003){Zoccali}, {Renzini}, {Ortolani}, {Greggio},
  {Saviane}, {Cassisi}, {Rejkuba}, {Barbuy}, {Rich}, \& {Bica}}]{zocc+2003}
{Zoccali}, M., {Renzini}, A., {Ortolani}, S., {et~al.} 2003, \aap, 399, 931

\end{thebibliography}

\end{document}